\documentclass[12pt]{article}
\pdfoutput=1
\usepackage[a4paper]{geometry}
 
\usepackage[T1]{fontenc} 
\usepackage{comment}

\usepackage{jheppub,hyperref,float,array,adjustbox,mathtools,physics,xcolor}
\usepackage{lipsum}
\usepackage{booktabs}
\usepackage{pdflscape}
\usepackage{geometry}
\usepackage{fancyhdr}
\usepackage{changepage}
\usepackage[english]{babel}

\usepackage{xcolor,tikz,pgfplots,amsmath,amsfonts}
\usepackage{graphicx}
\usepackage[mathscr]{euscript}
\usetikzlibrary{matrix,calc,positioning,decorations.markings,decorations.pathmorphing,decorations.pathreplacing}
\usetikzlibrary{arrows,cd}
\usetikzlibrary{shapes.misc}
\usetikzlibrary {shapes.geometric}
\usetikzlibrary {shapes.multipart}
\usetikzlibrary{decorations.markings}
\usetikzlibrary{backgrounds}
\usetikzlibrary{shapes.gates.logic.US}
\usetikzlibrary{circuits.ee.IEC}
\usepackage{bbding}

\definecolor{terradisiena}{RGB}{233,116,81}
\definecolor{strisciadipietro}{RGB}{229,204,255}
\definecolor{verdepetrolio}{RGB}{33,100,119}

\usepackage{tikz}
\usetikzlibrary{automata,positioning,calc}
\usetikzlibrary{decorations.markings}
\tikzset{->-/.style={decoration={markings, mark=at position #1 with {\arrow{>}}},postaction={decorate}}}
\tikzset{-<-/.style={decoration={markings, mark=at position #1 with {\arrow{<}}},postaction={decorate}}}
\tikzset{auto shift/.style={auto=right,->, to path={ let \p1=(\tikztostart),\p2=(\tikztotarget), \n1={atan2(\y2-\y1,\x2-\x1)},\n2={\n1+180} in ($(\tikztostart.{\n1})!1mm!270:(\tikztotarget.{\n2})$) -- ($(\tikztotarget.{\n2})!1mm!90:(\tikztostart.{\n1})$) \tikztonodes}}}

\usetikzlibrary {shapes.geometric}

\DeclareMathOperator{\USp}{USp}
\DeclareMathOperator{\SO}{SO}
\DeclareMathOperator{\SU}{SU}
\DeclareMathOperator{\UU}{U}

\newcommand{\ee}{\mathrm{e}}
\newcommand{\mi}{\mathrm{i}}

\newcommand{\asymmF}
{
	\begin{array}{c}
		\vspace{-2.8mm}
		\square \\
		\square
	\end{array}
}
 
\newcommand{\asymmBF}
{
	\begin{array}{c}
		\vspace{-2.8mm}
		\overline
		\square \\
		\square
	\end{array}
}

\newcommand{\symmF}
{ \square\!\square}

\newcommand{\symmBF}
{ \overline{\square\!\square}}

\definecolor{USPcol}{rgb}{0.94, 0.73, 0.80}
\definecolor{SUcol}{rgb}{0.894, 0.902, 0.976}
\definecolor{SOcol}{rgb}{0.5, 0.85, 0.8}

\title{
\begin{center}
4d/3d reduction of dualities with O6  
\end{center}
}

\author[a]{Antonio Amariti,}	
\author[a,b]{Pietro Glorioso,}	
\author[a,b]{Chiara Mascherpa,}
\author[a,b]{and Andrea Zanetti}

\affiliation[a]{INFN, Sezione di Milano, Via Celoria 16, I-20133 Milano, Italy}
\affiliation[b]{Dipartimento di Fisica, Università degli studi di Milano, Via Celoria 16, I-20133}

\emailAdd{antonio.amariti@mi.infn.it}
\emailAdd{pietro.glorioso@mi.infn.it}
\emailAdd{chiara.mascherpa@mi.infn.it}
\emailAdd{andrea.zanetti@mi.infn.it}

\abstract{
We consider $\mathrm{U}(N)$ gauge theories with a pair of two-index tensors interacting through a quartic superpotential, in addition to fundamentals and antifundamentals.
The models have a brane engineering  in terms of NS, D4, D6 branes and an O6 plane.
Depending on the representation of the tensorial matter we have either an O6$^{+}$ plane, an  O6$^{-}$ plane
or a combined state of O6$^{+}$  and  O6$^{-}$,
with the  addition of 8 semi-infinite half-D6 branes, where the  last  case realizes a chiral theory.
The 4d IR duality is realized through an HW transition in the brane description.
Here we study the circle reduction of these dualities from the brane 
perspective by T-dualizing along the compact direction.
We then compare the results against the one obtained from field theoretical considerations and from localization, finding a precise agreement. When we consider the reduction of the 4d superconformal index to the 3d squashed three sphere partition function we observe that it is not always possible to obtain convergent 3d result with the standard reduction prescription, and that the double scaling limit is necessary.}

\begin{document}
\maketitle
\flushbottom
\allowdisplaybreaks

\section{Introduction}
\label{sec:intro}

The circle reduction of 4d supersymmetric QFTs has been a fertile field of research in the last decade.
The main motivation beyond such studies concerns the fate of 4d dualities \cite{Seiberg:1994pq,Intriligator:1995id,Intriligator:1995ne} and their connection with 3d analogous dualities \cite{Aharony:1997bx,Aharony:1997gp,deBoer:1997kr,Intriligator:1996ex,Karch:1997ux}.
Such field of research has been boosted by supersymmetric localization, because supersymmetric partition functions on curved space 
are necessarily equivalent among the dual phases.
Such equivalences correspond to non trivial integral identities, in some case already known to the mathematical community
(see e.g. \cite{,Rains2009LimitsOE, Rains2010TransformationsOE}). 
The problem of finding 3d dualities starting from 4d ones translates in this language to computing opportune limits of the corresponding 4d
identities and deriving the  relative  3d ones. This is a necessary and quite strong  requirement in order to claim the existence of the relative 3d duality.

In the cases with four supercharges, the prescription that has allowed the derivation of myriads of 3d dualities starting from 4d ones has been developed in \cite{Aharony:2013dha}
and we will denote it as the ARSW (Aharony-Razamat-Seiberg-Willett) prescription in the rest of the paper.
It consists of considering the circle reduction of the 4d models by keeping the finite size effects due to the circle. Such effects are constraints on the compact Coulomb branch that is generated in the
reduction. Namely a superpotential for the Kaluza-Klein (KK) monopole (see e.g. \cite{Davies:1999uw,Davies:2000nw}) has to be added so as to constrain the symmetries of the 3d theory, 
preventing the generation of axial and topological symmetries and thereby preserving the same  global symmetry structure of the  parent 4d dualities.
It follows that the addition of this superpotential allows to maintain the 4d duality, giving rise to an effective 3d duality on the circle. Removing the effects of the KK monopole is possible by considering  opportune real mass deformation. Pure 3d models are obtained once the massive fields are integrated out.

The prescription has a localization counterpart in terms of the reduction of the supersymmetric partition function evaluated on $S^3 \times S^1$ to the partition function evaluated on the squashed three sphere $S^3$.

Up to an overall contribution corresponding to the supersymmetric Casimir energy \cite{Assel:2014paa,Assel:2015nca}, the 4d partition function  coincides with the 4d superconformal index \cite{Kinney:2005ej,Romelsberger:2005eg}.
This is a function of the fugacities associated to the $\SO(4) = \SU(2)_L \times \SU(2)_R$  isometries of $S^3$, denoted as $p$ and $q$ in the literature and of the flavor fugacity, 
that we denote as $v_k$ (where the index $k$ runs over the Cartan of the flavor symmetry).
We then refer to the 4d  index as $I(p,q;v_k)$, such that the 4d duality between an electric and a magnetic gauge theory corresponds to the relation 
\begin{equation}
\label{rel:4dindex}
I_{\text{ele}} (p,q;v_k) =I_{\text{mag}} (p,q;v_k) \,.
\end{equation}

The ARSW prescription corresponds to the limit $p,q,v_k \rightarrow 1$; namely it corresponds to finding the leading saddle-point approximation to the two sides of (\ref{rel:4dindex})  as $p,q \rightarrow 1$ and it gives rise to 
\begin{equation}
\label{rel:4dindexto3d}
\begin{array}{ccc}
I_{\text{ele}} (p,q;v_k) & = &I_{\text{mag}} (p,q;v_k) \\
 \downarrow &  &\downarrow  \\
f(\beta) Z_{\text{ele}} (\omega_1 ,\omega_2 ;m_k) & = & f(\beta) Z_{\text{mag}} (\omega_1 ,\omega_2 ;m_k) \,,\\
\end{array}
\end{equation}
where the divergent factors are absorbed in the function $f(\beta)$. The final identity 
$ Z_{\text{ele}} (\omega_1 ,\omega_2 ;m_k) = Z_{\text{mag}} (\omega_1 ,\omega_2 ;m_k) $
 can be interpreted as equality of two supersymmetric partition functions on the squashed three sphere, denoted here as $Z(\omega_1 ,\omega_2 ;m_k)$, where $\omega_{1,2}$ are the squashing parameters and
 $m_k$ correspond to the real masses.
 
 Identities of this type, corresponding to the leading saddles of the two sides of (\ref{rel:4dindex}) as $p, q \rightarrow 1$, are valid provided the constraints among the flavor fugacities of the 4d index translate into constraints among the real masses $m_k$. The constraints (denoted in the mathematical literature as balancing conditions) arise from consistency in 4d. Namely, they forbid the generation of topological and anomalous axial symmetries, and therefore correspond to the constraints imposed in the effective  3d theories on $S^1$  by the KK monopole superpotential.
 
 Summarizing, the leading saddle that gives rise to the identity in the last line of (\ref{rel:4dindexto3d}) 
corresponds to the application of the ARSW prescription at the level of the reduction (see also \cite{Dolan:2011rp,Gadde:2011ia,Imamura:2011uw,Niarchos:2012ah}) of the superconformal index to the three sphere partition function, round  \cite{Kapustin:2009kz,Jafferis:2010un,Hama:2010av} or squashed \cite{Hama:2011ea}. 
This is also the starting point to  deduce various dualities by removing the finite size effects through real mass flow.

In the last decade, starting from the analysis of \cite{Choi:2018hmj}, we witnessed to extensive studies of the $p,q\rightarrow 1$ limit of the superconformal index, thought as a matrix integral over the gauge holonomies, generically focusing beyond the dominant saddle point which is associated to the case of unbroken gauge symmetry. Such sub-leading saddles are generically found  by considering  different scalings for the flavor holonomies $v_k$. 
While originally relevant for the study of dual gravitational states,
the pattern of gauge symmetry breaking in such saddles potentially yields to new 3d dualities when (\ref{rel:4dindex}) is considered and such dualities  necessitate to be studied independently.
 
In \cite{Amariti:2024bdd} the approach of looking at sub-leading saddles of  the $p,q \rightarrow 1$ limit of (\ref{rel:4dindex}) was referred to as double scaling limit (see also \cite{ArabiArdehali:2015ybk,Hwang:2018riu} for related studies in the case of $\mathcal{N}=1$ models and 
\cite{Deb:2025ypl} for recent discussion in the $\mathcal{N}=2$ case).  
The reason for using such terminology is that we consider a double scaling on the radius of the circle and on the real masses that arise from the global symmetries in the compactification.
Such double scaling has been shown to play a crucial role in the analysis of the reduction of the duality for orthogonal SQCD, where the ARSW prescription cannot be applied to the three sphere partition function, because the effective duality on $S^1$ leaves an unlifted Coulomb branch that reflects into a divergent partition function \cite{Aharony:2013kma}.

If one deals with a duality that can be engineered in an Hanany-Witten (HW) setup \cite{Hanany:1996ie},  the double scaling just discussed at the level of the reduction of the superconformal admits a very natural interpretation at the level of brane picture.
The prescription in this case was discussed in \cite{Amariti:2015yea,Amariti:2015mva,Amariti:2016kat}
and it consists of T-dualizing the compact  space direction. 
The brane picture is quite useful when one looks for the 3d dual saddles (either leading or sub-leading).
Indeed, if one considers a saddle in the $p,q \rightarrow 1$ limit of the electric index, finding the dual one requires the correct guess of the gauge symmetry breaking pattern of the magnetic description.
A necessary (and often sufficient) condition in order to find such a guess consists of finding a dual pattern that cancels the divergent terms encoded in the function $f(\beta)$ in (\ref{rel:4dindexto3d}). 

On the other hand, the brane picture offers a direct approach in order to find such a dual pattern, consisting of the application of the HW transition, i.e. exchanging the NS branes in the T-dual setup.
As discussed extensively in \cite{Amariti:2015mva} this approach becomes quite useful when two-index tensorial matter is considered, because in such cases the presence of orientifolds carrying RR charges makes the T-dual setups quite intricate. This is because the O-planes (and their RR charges) split in pairs under T-duality, and this splitting plays a non-trivial role in the HW transition.

Motivated by these complexities and the insights offered by the double scaling approach, in this paper we consider a configuration that has not been studied  in full details in the literature of the 4d/3d reduction, consisting of  brane setups  with  O6 planes \cite{Landsteiner:1997vd,Landsteiner:1997ei,Csaki:1998mx,Landsteiner:1998gh,Brunner:1998jr,Elitzur:1998ju}.
The models under investigation have an O6 placed on an NS (or NS$'$) brane and three configurations are possible: beyond the standard  O6$^+$ and O6$^-$ configurations, a  mixed configuration  can also occur. This mixed  configuration corresponds to a split of the orientifold on an NS$'$ brane into two  semi-infinite O6 planes, with opposite RR charge. In this case RR charge conservation requires to add on the O6$^{-}$  plane eight half-D6 branes, semi-infinite in the direction associated to the FI coupling in the brane picture \cite{Brodie:1997sz,Hanany:1997sa}. This configuration allows for descriptions of dualities among chiral gauge theories. 

The cases with an O6$^{+}$ or an O6$^{-}$ realize a $\UU(N)$ model with a conjugate pair\footnote{In the following we will denote conjugated pairs of representations as a flavor.}  of two-index symmetric or antisymmetric tensors respectively.
On the other hand, the model with the combined O6$^+$/ O6$^-$ planes  allows to engineer models  with a conjugate symmetric tensor, an antisymmetric tensor, eight fundamentals and $F$  fundamental flavors. 
In each case an IR Seiberg-like duality has been studied in the literature through field theoretical considerations \cite{Intriligator:1995ax,Brodie:1996xm} and confirmed by  the brane setup \cite{Landsteiner:1997vd,Landsteiner:1997ei,Csaki:1998mx,Landsteiner:1998gh,Brunner:1998jr,Elitzur:1998ju}.

Observe that the duality for the setup with a conjugated symmetric and an antisymmetric 
was originally studied in \cite{Brodie:1996xm} by considering only the  superpotential involving the conjugated symmetric and the antisymmetric, necessary in order to guarantee the truncation of the chiral ring. Actually this duality does not have a direct description in terms of branes, but it is possible to find an engineering in presence of further superpotential deformations \cite{Landsteiner:1998gh,Brunner:1998jr,Elitzur:1998ju}. Furthermore, by triggering opportune mass flows (and consistently Higgs flows in the dual case)  one can flow back to the duality of \cite{Brodie:1996xm}.

In this paper we aim to study the circle reduction of the corresponding dualities for the three types of O6 planes discussed above. 
While the case with an O6$^+$ plane can be studied through the original ARSW prescription, we observe that the other two cases require a more sophisticated analysis, where the double scaling plays a crucial role. 
We then start our analysis by studying the case with the O6$^+$ both using the ARSW prescription and the double scaling limit, finding precise agreement.

Furthermore we exploit the double scaling formalism, at the level of the 
three sphere partition function, to predict the pure 3d limit in a third different way that reveals crucial in the analysis of the cases with O6$^{-}$ planes.
Namely, the symmetry structure of the models on $S^1$, marked by the fact that in the T-dual configurations the orientifolds split symmetrically on the circle, allows one to find an identity between an electric and a magnetic partition function with the structure $Z_{\text{ele}}^2 = Z_{\text{mag}}^2$. This identity is obtained by opportunely identifying the mass parameters and we observe in the case with the O6$^+$ that the final results coincides with the ones found by the other two methods, after extracting the square root of the relation.
This last formalism can be used also in the cases with the O6$^-$ plane, where the ARSW prescription cannot be applied to the reduction of the superconformal index to the three sphere partition function.
This allows us to obtain the pure 3d limit in all the cases, finding compatible results from the field theory analysis, from the brane picture and from the localization approach.

The paper is organized as follows. In Section \ref{sec:rev}, we survey 4d dualities involving two-index tensors, which we aim to reduce to 3d. We first present their $\SU(N)$ version, followed by a discussion of their $\UU(N)$ counterpart and the conjectural identities at the level of the superconformal index. Section \ref{sec:branes} studies these dualities through type II brane pictures. We review existing type IIA brane setups before introducing the type IIB configurations obtained via compactification along the $x_{3}$ direction and T-duality. This analysis yields the effective and pure 3d dualities from a brane perspective. In Section \ref{sec:genscaling}, we explore the field theory reduction using the double scaling limit of the superconformal index, successfully reproducing the results from the brane picture analysis. Section \ref{sec:end} surveys our main results and discusses future lines of research.
The appendices provide supplementary details: Appendix \ref{dsrev} collects formulas for computing the double scaling limit of the squashed three-sphere partition function.  Appendices \ref{app:0} and \ref{app:U4}  discuss two confining dualities that played a crucial role in extracting the superpotentials of the 3d limits. Indeed  they give origin to the electric monopoles acting as singlets in the dual phase, a hallmark of 3d dualities with vanishing Chern-Simons levels. Finally, Appendix \ref{app:mono} comments on monopoles and branching rules in the presence of two-index symmetric and antisymmetric tensors.

\section{A survey of the 4d dualities}
\label{sec:rev}

In this section we collect the 4d dualities that we wish to reduce to 3d in the 
bulk of the paper. We start by providing the field content, the superpotential and the duality map through the global charge structure for the electric and the magnetic phase. We then provide the relative identity between the superconformal indices. We stress here that such identities are conjectural, i.e. they have not been proven 
so far in the literature (only perturbative derivations and limiting cases have be checked). 
A proof of the identities (e.g. by exploiting the tensor deconfinement technique, as recently done in \cite{Benvenuti:2024glr,Hwang:2024hhy} for the KSS duality \cite{Kutasov:1995ss}) is beyond the scopes of this paper.
Here we  conjecture the validity of such identities and we show  that the matching of the 3d squashed three sphere partition function follows from the 
matching of the 4d indices.

\subsection{4d duality  for $\SU(N)$ with $\tilde A$ and $A$}

The first 4d duality that we aim  to reduce to 3d was originally found in \cite{Intriligator:1995ax}.
The electric theory has an $\SU(N)$ gauge group with a conjugate  antisymmetric
tensor $\tilde A$, an antisymmetric $A$, $F$ fundamentals $Q$ and antifundamentals $\tilde Q$. In addition to such field content there is a superpotential 
\begin{equation}
\label{WAAt}
W  = (A \tilde A)^2.
\end{equation}
This theory can be generalized by considering an adjoint $\Phi$ and superpotential
\begin{equation}
\label{WAAtphi}
W  = \Phi^{k+1}+ A \Phi \tilde A\,.
\end{equation}
The superpotential (\ref{WAAtphi}) reduces to (\ref{WAAt}) for  $k=1$.
Further deformations of this model consist of adding interactions like $(Q \tilde Q)^2$ and $Q A \tilde A \tilde Q$.

In absence of such deformations the dual model corresponds to an $\SU(\tilde N)$ 
gauge theory with $\tilde N = 3F-N-4$, a conjugate  antisymmetric
tensor $\tilde a$, an antisymmetric $a$, $F$ dual fundamentals $q$ and antifundamentals $\tilde q$.
In addition there are mesonic flippers $M_0=Q \tilde Q$, $M_1 = Q A \tilde A \tilde Q$, $P= \tilde A Q^2$ and $\tilde P= A\tilde Q^2$, such that the dual superpotential becomes
\begin{equation}
\label{Wdualaat}
W = (a \tilde a)^2 + M_0 q a \tilde a q + M_1 q \tilde q+ P \tilde a q^2 +\tilde P a
\tilde  q^2 .
\end{equation}
The global charges of the electric theory are summarized in the table below.
\begin{equation}
\begin{array}{c|c|ccccc}
               &\SU(N)&\SU(F)   & \SU(F) & \UU(1)_B & \UU(1)_X & \UU(1)_R\\
               \hline
Q            & \square &  \square &1 &   \frac{1}{N}&0 &1- \frac{N+2}{2F} \\
\tilde Q    & \overline \square & 1  &  \square &\! \! \! \! \! -\frac{1}{N} &0  & 1- \frac{N+2}{2F} \\
\tilde A    & \begin{array}{c} \overline \square \vspace{-2.9mm} \\  \square  \end{array}  &1   & 1&\! \! \! \! \!   -\frac{2}{N}&\! \! \! \! \! -1  &\frac{1}{2} \\
A            & \begin{array}{c} \square \vspace{-2.9mm} \\  \square  \end{array} &1    & 1&\frac{2}{N} & 1 &\frac{1}{2}  \\
\end{array}
\end{equation}
The global symmetry group of the magnetic theory coincides with the one of the electric description and the fields are charged as showed below.
\begin{equation}
\begin{array}{c|c|cccccc}
&\SU(\tilde N)&\SU(F)   & \SU(F) & \UU(1)_B & \UU(1)_X & \UU(1)_R\\
\hline
q            & \square &  \overline \square &1 &   \frac{1}{\tilde N}& \frac{F-2}{\tilde N} &1- \frac{\tilde N+2}{2F} \\
\tilde q    & \overline \square & 1  &  \overline\square&\! \! \! \! \! -\frac{1}{\tilde N} &\! \! \! \! \!   -\frac{F-2}{\tilde N} & 1- \frac{\tilde N+2}{2F} \\
\tilde a    &\begin{array}{c} \overline \square \vspace{-2.9mm} \\  \square  \end{array}  & 1& 1&\! \! \! \! \!  -\frac{2}{\tilde N}&\! \! \! \! \!  - \frac{N-F}{\tilde N}&\frac{1}{2} \\
a            & \begin{array}{c} \square \vspace{-2.9mm} \\  \square  \end{array}    &1 & 1&\frac{2}{\tilde N} &  \frac{N-F}{\tilde N}&\frac{1}{2}  \\
\hline 
P             & 1  & \begin{array}{c} \square \vspace{-2.9mm} \\  \square  \end{array}  & 1& 0  &\! \! \! \! \! -1 &\frac{5}{2}- \frac{N+2}{F} \\
\tilde P    & 1  & 1  & \begin{array}{c} \square \vspace{-2.9mm} \\  \square  \end{array}  &0   &1 &\frac{5}{2}- \frac{N+2}{F}  \\
M_0            & 1  & \square  &\square  & 0 & 0&  2- \frac{N+2}{F} \\
N _1           &  1  & \square  & \square & 0 & 0& 3- \frac{N+2}{F}\\
\end{array}
\end{equation}
The duality among the $\SU(N)$ and $\SU(\tilde N)$ theories can be traded with a duality among $\UU(N)$ and $\UU(\tilde N)$ by gauging the baryonic symmetry. In the following we will consider such a duality because it is the one that is actually observed from the brane picture, where the abelian factor is  realized explicitly.

We can also read the integral formulas associated to the superconformal index of the electric and of the magnetic phase. We refer the reader to \cite{Spiridonov:2009za} for the identity between the models without the further $\UU(1)_B$ gauging.
In the electric case we have\footnote{When considering the reduction to 3d we will also have an  FI parameter, added to the index following the discussion of \cite{Aharony:2013dha}.}
\begin{equation}
\label{indexele0}
I_{\UU(N)}^{[F\square;F \overline \square; 1  \overline{A};1A]} 
(\vec \mu;\vec \nu; t_{\tilde A};t_A)\,,
\end{equation}
where the arguments are separated by a semicolon if they transform under  different representations. The  representations are specified in the square brackets   and 
we further parametrize the fugacities in the argument of $I_{\UU(N)}$ as
\begin{equation}
\label{parele0}
\begin{array}{lll}
\mu_b = m_b (pq)^{\frac{1}{2} \left(1-\frac{N+2}{2F}\right)}, & \quad b=1,\dots,F\,, &\quad
\text{with} \quad \prod_{b=1}^F m_b=1\,,\\
\nu_b = n_b (pq)^{\frac{1}{2} \left(1-\frac{N+2}{2F}\right)}, &\quad b=1,\dots,F\,, &\quad
\text{with} \quad \prod_{b=1}^F n_b=1\,,\\
t_{\tilde A} = x^{-1} (pq)^{\frac{1}{4}}, \\
t_A = x (pq)^{\frac{1}{4}} .\\
\end{array}
\end{equation}
The balancing condition of \cite{Spiridonov:2009za}, namely
\begin{equation}
\prod_{b=1}^F \mu_b \,\nu_b
=
(pq)^{F-\frac{N}{2}-1},
\end{equation} 
is automatically satisfied by our parametrization.
The magnetic index is
\begin{equation}
\prod_{1 \leq b < c \leq F} \Gamma_e(t_{\tilde A} \mu_b \mu_c,
t_A \nu_b \nu_c)
\prod_{b,c=1}^F \Gamma_e(\mu_b \nu_c; \mu_b \nu_c t_A t_{\tilde A})\;
I_{\UU(\tilde N)}^{[F\square;F \overline \square; 
1 \overline A;1 A]} 
(\vec {\tilde \mu};\vec {\tilde \nu};  \tilde t_{\tilde a};\tilde t_a),
\end{equation}
where the duality dictionary translates into the 
parametrization of the dual fugacities in the argument of $I_{\UU(\tilde N)}$ as
\begin{equation}
\begin{array}{l}
\tilde \mu_b= m_b^{-1} x^{\frac{F-2}{\tilde N}} (pq)^{\frac{1}{2} \left(1-\frac{\tilde N+2}{2F}\right)}
,\\
\tilde  \nu_b = n_b^{-1} x^{-\frac{F-2}{\tilde N}} (pq)^{\frac{1}{2} \left(1-\frac{\tilde N+2}{2F}\right)}
, \\
t_{\tilde a} =  x^{-\frac{N-F}{\tilde N}}  (pq)^{\frac{1}{4}} 
,\\
t_a = x^{\frac{N-F}{\tilde N}}  (pq)^{\frac{1}{4}} ,
\end{array}
\end{equation}
where  $b=1,\dots,F$ as in \eqref{parele0}.

\subsection{4d duality  for $\SU(N)$ with $\tilde S$ and $S$}

The second 4d duality that we aim  to reduce to 3d was originally found in \cite{Intriligator:1995ax}.
The electric theory has an $\SU(N)$ gauge group with a conjugate  symmetric
tensor $\tilde S$, a symmetric tensor $S$, $F$ fundamentals $Q$ and antifundamentals $\tilde Q$. In addition to such field content there is a superpotential 
\begin{equation}
\label{WSSt}
W  = (S \tilde S)^2.
\end{equation}
This theory can be generalized by considering an adjoint $\Phi$ and superpotential
\begin{equation}
\label{WSStphi}
W  = \Phi^{k+1}+ S \Phi \tilde S,
\end{equation}
where the superpotential (\ref{WSStphi}) reduces to (\ref{WSSt}) for  $k=1$.
Further deformations of this model consist of adding interactions like $(Q \tilde Q)^2$ and $Q S \tilde S \tilde Q$.

In absence of such deformations the dual model corresponds to an $\SU(\tilde N)$ 
gauge theory with $\tilde N = 3F-N+4$, a conjugate  symmetric
tensor $\tilde s$, a symmetric tensor $s$, $F$ dual fundamentals $q$ and antifundamentals $\tilde q$.
In addition there are mesonic flippers $M_0= Q \tilde Q$, $M_1 = Q S \tilde S \tilde Q$, $P= \tilde  S Q^2$ and $\tilde P= S\tilde Q^2$, such that the dual superpotential becomes
\begin{equation}
\label{Wdualsst}
W = (s \tilde s)^2 + M_0 q s \tilde s q + M_1 q \tilde q+ P \tilde s q^2 +\tilde P s
\tilde  q^2 .
\end{equation}
The global charges of the electric theory are summarized in the table below.
\begin{equation}
\begin{array}{c|c|ccccc}
               &\SU(N)&\SU(F)   & \SU(F) & \UU(1)_B & \UU(1)_X & \UU(1)_R\\
               \hline
Q            & \square &  \square &1 &   \frac{1}{N}&0 &1- \frac{N-2}{2F} \\
\tilde Q    & \overline \square & 1  &  \square &\! \! \! \! \! -\frac{1}{N} &0  & 1- \frac{N-2}{2F} \\
\tilde S    & \square \!\square  &1   & 1& \! \! \! \! \!  -\frac{2}{N}&\! \! \! \! \! -1  &\frac{1}{2} \\
S            & \overline{\square \!\square} &1    & 1&\frac{2}{N} & 1 &\frac{1}{2}  \\
\end{array}
\end{equation}
The global symmetry group coincides with the one of the electric description and the fields are charged as showed below.
\begin{equation}
\begin{array}{c|c|cccccc}
               &\SU(\tilde N)&\SU(F)   & \SU(F) & \UU(1)_B & \UU(1)_X & \UU(1)_R\\
               \hline
q            & \square &  \overline \square &1 &   \frac{1}{\tilde N}& \frac{F+2}{\tilde N} &1- \frac{\tilde N-2}{2F} \\
\tilde q    & \overline \square & 1  &  \overline\square&\! \! \! \! \! -\frac{1}{\tilde N} &\! \! \! \! \!   -\frac{F+2}{\tilde N} & 1- \frac{\tilde N-2}{2F} \\
\tilde s    &\overline{\square \! \square} & 1& 1&\! \! \! \! \!  -\frac{2}{\tilde N}&\! \! \! \! \!  - \frac{N-F}{\tilde N}&\frac{1}{2} \\
s           & \square \! \square   &1 & 1&\frac{2}{\tilde N} &  \frac{N-F}{\tilde N}&\frac{1}{2}  \\
\hline 
P             & 1  & \square \! \square & 1& 1  &\! \! \! \! \! -1 &\frac{5}{2}- \frac{N-2}{F} \\
\tilde P    & 1  & 1  &  \square \! \square  &1   &1 &\frac{5}{2}- \frac{N-2}{F}  \\
M_0            & 1  & \square  &\square  & 1 & 0&  2- \frac{N-2}{F} \\
N _1           &  1  & \square  & \square & 1 & 0& 3- \frac{N-2}{F}\\
\end{array}
\end{equation}
The duality among the $\SU(N)$ and $\SU(\tilde N)$ theories can be traded with a duality among $\UU(N)$ and $\UU(\tilde N)$ by gauging the baryonic symmetry. We will once again consider such a duality because it is the one observed in the brane picture, where the abelian factor is  realized explicitly.

We can also read the integral formulas associated to the superconformal index of the electric and of the magnetic phase. We refer the reader to \cite{Spiridonov:2009za} for the identity between the models without the further $\UU(1)_B$ gauging.
In the electric case we have 
\begin{equation}
\label{indexeleS}
I_{\UU(N)}^{[F \square;F \overline \square; 1 \overline{S},1S]} 
(\vec \mu;\vec \nu; t_{\tilde S};t_S)\,,
\end{equation}
where, again, the arguments are separated by a semicolon if they transform under a different representations. The  representations are specified in the square brackets   and 
we further parametrize the fugacities in the argument of $I_{\UU(N)}$ as
\begin{equation}
\label{pareleS}
\begin{array}{lll}
\mu_b = m_b (pq)^{\frac{1}{2} \left(1-\frac{N-2}{2F}\right)}, & \quad b=1,\dots,F\,, &\quad
\text{with} \quad \prod_{b=1}^F m_b=1\,,\\
\nu_b = n_b (pq)^{\frac{1}{2} \left(1-\frac{N-2}{2F}\right)}, &\quad b=1,\dots,F\,, &\quad
\text{with} \quad \prod_{b=1}^F n_b=1\,, \\
t_{\tilde S} = x^{-1} (pq)^{\frac{1}{4}},\\
t_S = x (pq)^{\frac{1}{4}}. \\
\end{array}
\end{equation}
The balancing condition of \cite{Spiridonov:2009za}, namely
\begin{equation}
\label{BCSSt4d}
\prod_{b=1}^F\mu_b \, \nu_b
=
(pq)^{F-\frac{N}{2}+1},
\end{equation} 
is automatically satisfied by our parametrization.
The magnetic index is
\begin{equation}
\prod_{1 \leq c \leq  b \leq F} \Gamma_e(t_{\tilde S} \mu_c \mu_b,t_S \nu_c \nu_b)
\prod_{a,b=1}^F \Gamma_e(\mu_c \nu_b; \mu_c \nu_b t_S t_{\tilde S})
\;
I_{\UU(\tilde N)}^{[F\square;F \overline \square; 
1  \overline{S}; 
1  S]} 
(\vec {\tilde \mu};\vec {\tilde \nu};  \tilde t_{\tilde s};\tilde t_s)\,.
\end{equation}
Here the duality dictionary translates into the 
parametrization of the dual fugacities in the argument of $I_{\UU(\tilde N)}$ as
\begin{equation}
\begin{array}{l}
\tilde \mu_b = m_b^{-1} x^{\frac{F+2}{\tilde N}} (pq)^{\frac{1}{2} \left(1-\frac{\tilde N-2}{2F}\right)},
\\
\tilde  \nu_b = n_b^{-1} x^{-\frac{F+2}{\tilde N}} (pq)^{\frac{1}{2} \left(1-\frac{\tilde N-2}{2F}\right)},
 \\
t_{\tilde s} =  x^{-\frac{N-F}{\tilde N}}  (pq)^{\frac{1}{4}} ,
\\
t_s = x^{\frac{N-F}{\tilde N}}  (pq)^{\frac{1}{4}} ,
\end{array}
\end{equation}
where the range for the labels $b$ for the fundamentals and the antifundamentals  are omitted since they are identical to the ones appearing in  \eqref{pareleS}.

\subsection{4d duality  for $\SU(N)$ with $\tilde S$ and $A$}
The last duality under inspection descends from a duality originally derived in \cite{Brodie:1996xm}, where the electric theory is an $\SU(N)$ gauge group with a conjugate  symmetric
tensor $\tilde S$, an antisymmetric tensor $A$, $F+8$ fundamentals $Q$, $F$ antifundamentals $\tilde Q$ and superpotential 
\begin{equation}
\label{W1}
W  = (A \tilde S)^2.
\end{equation}
This theory can be generalized by considering an adjoint $\Phi$ and superpotential
\begin{equation}
\label{W2}
W  = \Phi^{k+1}+ A \Phi \tilde S\,.
\end{equation}
The superpotential (\ref{W2}) reduces to (\ref{W1}) for  $k=1$.
Further deformations of this model consist of adding interactions like $\tilde S Q^2$ and $A \tilde Q^2$ breaking the flavor symmetry to 
$\SO(2F+8)$ and $\USp(2F)$ respectively.

In the following we will mostly focus on the case with a superpotential deformation between the conjugate symmetric and eight out of the $F+8$ fundamentals, that we will denote as $T$.
The electric superpotential in this case is given by
\begin{equation}
\label{W1def}
W  = (A \tilde S)^2 + \tilde S T^2.
\end{equation}
On the other hand, the other $F$ fundamentals $Q$ and $F$ antifundamentals $\tilde Q$ are not involved in the superpotential.
This is the electric configuration studied in detail in \cite{Landsteiner:1998gh}. In the following we will review the details of the duality necessary to our analysis.
The global symmetry group is $\SU(F)^2 \times \SO(8) \times \UU(1)_B \times \UU(1)_X \times \UU(1)_R$ and the fields are charged as showed in the table below.
\begin{equation}
\begin{array}{c|c|cccccc}
               &\SU(N)&\SU(F)   & \SU(F) &\SO(8) & \UU(1)_B & \UU(1)_X & \UU(1)_R\\
               \hline
Q            & \square &  \square &1 & 1&\frac{1}{N}&0 &1- \frac{N-2}{2F} \\
\tilde Q    & \overline \square & 1  &  \square&1 &\! \! \! \! \! -\frac{1}{N} &0  & 1- \frac{N-2}{2F} \\
T            & \square &  1&1 & 8_V& \frac{1}{N}& \frac{1}{2} &\frac{3}{4} \\
\tilde S    & \overline{\square\!\square} &1   & 1& 1&\! \! \! \! \!  -\frac{2}{N}&\! \! \! \! \! -1  &\frac{1}{2} \\
A            & \begin{array}{c} \square \vspace{-2.9mm} \\  \square  \end{array} &1   &1 & 1&\frac{2}{N} & 1 &\frac{1}{2}  \\
\end{array}
\end{equation}
The dual theory has an $\SU(\tilde N)$ gauge group with $\tilde N = 3F-N+4$, $F$ conjugated pairs of fundamentals and antifundamentals $q$ and $\tilde q$, eight extra fundamentals $t$, a conjugated symmetric $\tilde s$ and an antisymmetric $a$. There are also various mesons of the electric theory acting as singlets in the dual picture, denoted as $P = \tilde S Q^2$, $\tilde P = A \tilde Q^2$, $N = Q A\tilde S \tilde Q$, $M = Q \tilde Q$ and $L = T \tilde Q$.
The theory has superpotential 
\begin{equation}
W= (a \tilde s)^2+\tilde s t^2+ \tilde P \tilde q a \tilde q + P  q \tilde s  q + N \tilde q q + M q a \tilde s \tilde q+ L t a \tilde s \tilde q\,.
\end{equation}
The global symmetry group coincides with the one of the electric description and the fields are charged as showed below.
\begin{equation}
\begin{array}{c|c|cccccc}
               &\SU(\tilde N)&\SU(F)   & \SU(F) &\SO(8) & \UU(1)_B & \UU(1)_X & \UU(1)_R\\
               \hline
q            & \square &  \overline \square &1 & 1&  \frac{1}{\tilde N}& \frac{F+2}{\tilde N} &1- \frac{\tilde N-2}{2F} \\
\tilde q    & \overline \square & 1  &  \overline\square&1 &\! \! \! \! \! -\frac{1}{\tilde N} & \! \! \! \! \!  -\frac{F+2}{\tilde N} & 1- \frac{\tilde N-2}{2F} \\
t            & \square &  1&1 & 8_V& \frac{1}{\tilde N}&\frac{N-F}{2\tilde N} &\frac{3}{4} \\
\tilde s    & \overline{\square\!\square} &1   & 1& 1&\! \! \! \! \!  -\frac{2}{\tilde N}& \! \! \! \! \! - \frac{N-F}{\tilde N}&\frac{1}{2} \\
a            & \begin{array}{c} \square \vspace{-2.9mm} \\  \square  \end{array} &1   &1 & 1&\frac{2}{\tilde N} &  \frac{N-F}{\tilde N}&\frac{1}{2}  \\
\hline 
P             & 1  &\square\!\square  & 1& 1&0  &\! \! \! \! \! -1 &\frac{5}{2}- \frac{N-2}{F} \\
\tilde P    & 1  & 1  & \begin{array}{c} \square \vspace{-2.9mm} \\  \square  \end{array}  &1 &0  &1 &\frac{5}{2}- \frac{N-2}{F}  \\
M            & 1  & \square  &\square &1 & 0 & 0&  2- \frac{N-2}{F} \\
N            &  1  & \square  & \square&1 & 0 & 0& 3- \frac{N-2}{F}\\
L            &  1  &  1 &\square &8_V & 0 &\frac{1}{2} &\frac{7}{4}- \frac{N-2}{2F}  \\
\end{array}
\end{equation}
The duality among the $\SU(N)$ and $\SU(\tilde N)$ theories can be traded with a duality among $\UU(N)$ and $\UU(\tilde N)$ by gauging the baryonic symmetry. In the following we will consider such a duality because it is the one that is actually observed from the brane picture, where the abelian factor is indeed realized explicitly.

We can also read the integral formulas associated to the superconformal index of the electric and of the magnetic phase. We refer the reader to \cite{Spiridonov:2009za} for the identity between the models without the further $\UU(1)_B$ gauging and in absence of the superpotential deformation $\tilde S T^2$.
In the electric case we have 
\begin{equation}
\label{indexele}
I_{\UU(N)}^{[(F+8)\square;F \overline \square; 1 \overline{S};1A]} 
(\vec \mu,\vec w;\vec \nu; t_{\tilde S};t_A)\,,
\end{equation}
where the arguments are separated by a comma if they transform under  the same representation with respect to $\UU(N)$ and by a semicolon if they transform under a different representations. The  representations are specified in the square brackets   and 
we further parametrize the fugacities in the argument of $I_{\UU(N)}$ as
\begin{equation}
\label{parele}
\begin{array}{lll}
\mu_b = m_b (pq)^{\frac{1}{2} \left(1-\frac{N-2}{2F}\right)}, & \quad b=1,\dots,F\,, &\quad
\text{with} \quad \prod_{b=1}^F m_b=1\,,\\
\nu_b = n_b (pq)^{\frac{1}{2} \left(1-\frac{N-2}{2F}\right)}, &\quad b=1,\dots,F\,, &\quad
\text{with} \quad \prod_{b=1}^F n_b=1\,, \\
z_\ell =  w_\ell x^{\frac{1}{2}} (pq)^{\frac{3}{8}},& \quad \ell=1,\dots 4\,,\\
z_\ell = w_\ell^{-1}  x^{\frac{1}{2}}  (pq)^{\frac{3}{8}},& \quad   \ell=5,\dots 8\,, \\
t_{\tilde S} = x^{-1} (pq)^{\frac{1}{4}}, \\
t_A = x (pq)^{\frac{1}{4}}. \\
\end{array}
\end{equation}
Note that the balancing condition of \cite{Spiridonov:2009za}, namely
\begin{equation}
t_{\tilde S}^{N+2} t_A^{N-2} \prod_{b=1}^F \mu_b \, \nu_b  \prod_{\ell=1}^8 z_\ell  =(pq)^{F+4},
\end{equation} 
is automatically satisfied by our parametrization.
The magnetic index is
\begin{eqnarray}
&&
\prod_{1 \leq b \leq c \leq F} \Gamma_e(t_{\tilde S} \mu_b \mu_c)
\prod_{1 \leq b < c \leq F}\Gamma_e(t_A \nu_b \nu_c)
\prod_{b,c=1}^F \Gamma_e(\mu_b \nu_c; \mu_b \nu_c t_{\tilde S} t_A)
\prod_{b=1}^F\prod_{\ell=1}^8 \Gamma_e(\nu_b z_\ell)
\nonumber \\
&& \times \,
I_{\UU(\tilde N)}^{[(F+8)\square;F \overline \square; 1 \overline{S}; 1 A]} 
(\vec {\tilde \mu},\vec {\tilde w};\vec {\tilde \nu}; 	\tilde t_{\tilde s};\tilde t_a)\,.
\end{eqnarray}
Here the duality dictionary translates into the 
parametrization of the dual fugacities in the argument of $I_{\UU(\tilde N)}$ as
\begin{equation}
\begin{array}{l}
\tilde \mu_b = m_b^{-1} x^{\frac{F+2}{N}} (pq)^{\frac{1}{2} \left(1-\frac{\tilde N-2}{2F}\right)},
\\
\tilde  \nu_b = n_b^{-1} x^{-\frac{F+2}{N}} (pq)^{\frac{1}{2} \left(1-\frac{\tilde N-2}{2F}\right)},
 \\
\tilde  z_\ell =  w_\ell \,\,\,\,x^{\frac{N-F}{2\tilde N}} (pq)^{\frac{3}{8}},
\\
\tilde  z_\ell = w_\ell^{-1}  x^{\frac{N-F}{2\tilde N}}  (pq)^{\frac{3}{8}},
 \\
\tilde t_{\tilde s} =   x^{-\frac{N-F}{\tilde N}}  (pq)^{\frac{1}{4}} ,
\\
\tilde t_a = x^{\frac{N-F}{\tilde N}}  (pq)^{\frac{1}{4}} ,
\end{array}
\end{equation}
where the range for the labels $b$ for the fundamentals and the antifundamentals  are omitted since they are identical to the ones appearing in  \eqref{parele}.

\section{Branes}
\label{sec:branes}

In this section we provide the analysis for the reduction of the dualities from the brane perspective.
We refer the reader to \cite{Hanany:1996ie,Elitzur:1997fh,Elitzur:1997hc,Giveon:1998sr,Aharony:1997ju,deBoer:1998by,deBoer:1997ka,deBoer:1996ck} for general discussions on the brane engineering considered in this paper.
The case with a pair of conjugated  antisymmetric tensors has been analyzed in \cite{Amariti:2015mva}. The case with a pair of conjugated  symmetric tensors can be studied along the same lines, by considering an O6 plane with opposite RR charge.
For this reason in this section we will mostly focus on the chiral case, commenting on the other cases at the end of the section.

When we consider the duality with a conjugate symmetric and an antisymmetric 
we need to include  superpotential deformations, like the deformation $\tilde S T^2 $ discussed above, in order to have   an explicit realization in terms of HW setups.
The brane setup is realized by considering NS fivebranes, D4 and D6 branes and an orientifold six-plane. 

\begin{figure}
\begin{center}
  \includegraphics[width=14cm]{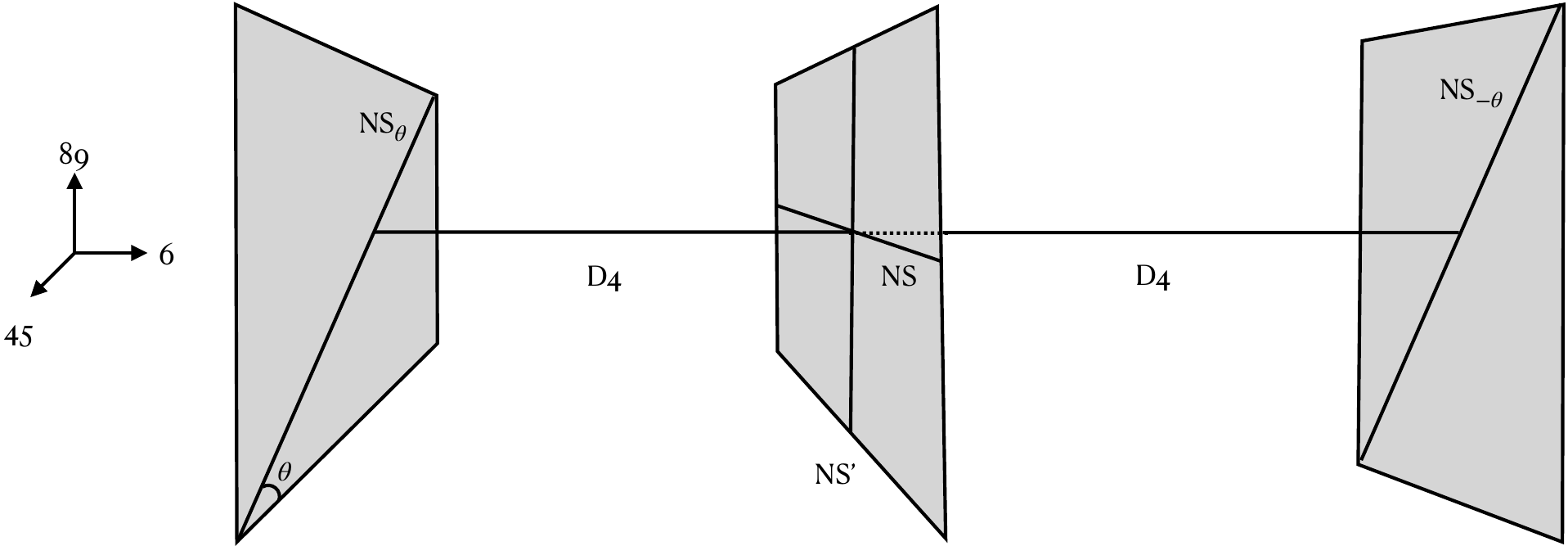}
  \end{center}
  \caption{In this figure we give a pictorial representation of the fivebranes and of the D4 branes considered in this paper, by representing  their extension in directions $x_{45689}$.}
    \label{Fig:NSD4}
\end{figure}

Following the usual conventions in the literature,  the fivebranes  are extended along the four spacetime directions $x_{0123}$ and two extra directions, corresponding to $x_{45}$ for an NS brane and $x_{89}$ for an NS$'$ brane.
When an NS brane is rotated in the  $x_{89}$ plane with respect to the $x_{45}$ plane by an angle $\theta$
we refer to such a brane as an NS$_{\theta}$.
The D4 branes are generically stretched between pairs of NS branes, placed at finite distance along the direction $x_6$.
We depicted the possible configurations of D4 and NS branes in Figure \ref{Fig:NSD4}.
Instead, the six-branes are denoted as D6 if they are extended along  $x_{0123789}$ and 
 D6$'$ if they are extended along  $x_{0123457}$.
Analogously to the case of the NS fivebrane, we can also rotate the D6 by an angle between the directions  $(4\,5)$ and $(8\,9)$, referring to such branes as D6$_\theta$.

\begin{figure}
\begin{center}
  \includegraphics[width=6cm]{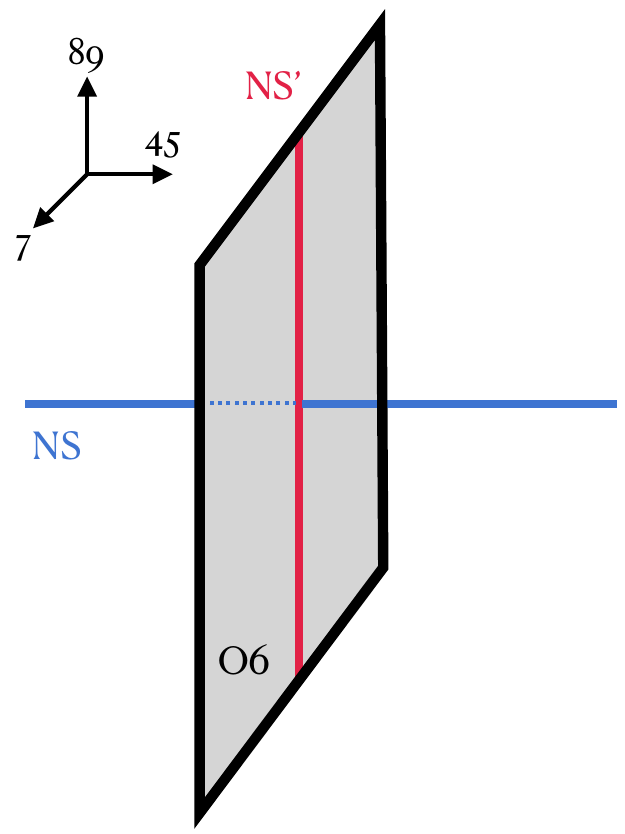}
  \end{center}
  \caption{In this figure we consider an O6 plane extended along the directions
   $x_{0123789}$, an NS brane extended along $x_{012345}$ and an NS$'$ brane $x_{012389}$. If we consider an O6 plane with an  NS$'$ brane the orientifold  changes its  RR charge at $x_7=0$ and RR  charge conservation imposes the addition of eight semi-infinite (along $x_7$) half-D6 branes.}
    \label{Fig:NSO6}
\end{figure}

The crucial ingredient, needed to realize the chiral gauge theories discussed above, is considering the configuration with an orientifold six-plane extended along $x_{0123789}$ with an NS$'$ brane (see Figure \ref{Fig:NSO6}).
We did not specify so far the projection on the RR sector of such O-plane, i.e. if we consider an O6$^+$ or an O6$^-$.
The reason is that, once we place an NS$'$ brane on the orientifold, the sign of the orientifold projection changes while crossing the fivebrane. Consistency with the RR charge conservation requires the addition on the O6$^-$ of eight semi-infinite (in the direction $x_7$) D6 branes.

We then consider a stack of $N$ D4  branes connecting an NS$_{\theta}$  and an NS$_{-\theta}$ fivebrane symmetrically with respect to the configuration with O6$^{\pm}$/NS$'$/8 half-D6 branes, in the direction $x_6$.
We therefore obtain an $\SU(N)$ gauge theory with an antisymmetric $ A$ a conjugate symmetric $\tilde S$ and eight fundamentals, with the superpotential 
discussed above in formula (\ref{W1def}).
In this setup we add extra $2F$ D6 branes, either D6$_{\pm \theta}$ of D6 or D6$'$. We have the following possibilities:

\begin{itemize}
\item  $\Delta W = 0$:  the flavor symmetry is $\SU(F)^2 \times \SO(8)$. In this case we have $F$ pairs of D6$_{\pm \theta}$ branes, placed symmetrically on the left and on the right of the O6/NS$'$/D6 stack. This is the models reviewed extensively in Section \ref{sec:rev}.
\item $\Delta W = \tilde S Q^2+A \tilde Q^2$: this deformation 
is realized by considering $2F$ D6 branes. In such case the extra D6 branes can be moved on the  O6/NS$'$/D6 stack, such that we have an $\SO(2 F+8) \times \USp(2F)$ flavor symmetry.
\item $\Delta W = (Q \tilde Q)^2+QA\tilde S \tilde Q$, this setup is realized by considering $2F$ D6$'$ branes, that, again, can be placed on the orientifold. The flavor symmetry in this case is $\SO(8) \times \SO(2F) \times \USp(2F)$.
\end{itemize}

Even if the discussion below can be generalized to $k>1$, in the following we restrict to the case $k=1$. The dual gauge group in absence of deformation is $\SU(3F-N+12)$. The deformations reflect into a dual Higgsings, such that the dual gauge group becomes $\SU(3F-N+4)$ in the first case and $\SU(2F-N+4)$
in the second and third one.
The dual gauge theory expectations are confirmed at brane level by moving the NS$_{\pm \theta}$ branes along $x_6$, exchanging their positions.
We refer the readers to the original references for details \cite{Landsteiner:1998gh,Brunner:1998jr,Elitzur:1998ju}.
Here for the ease of the reading we summarize the positions occupied by the various branes considered above in Table \ref{position} (see also \cite{Ahn:2007eh} for a generic picture representing  this brane setup).

\begin{table}[h!]
\centering
\begin{tabular}{l|cccccccccc}
              & 0 & 1 & 2 & 3 & 4 & 5 & 6 & 7 & 8 & 9 \\
\hline
D4            & X & X & X & X & $\cdot$  & $\cdot$ & X & $\cdot$ &  $\cdot$ & $\cdot$  \\
NS            & X & X & X & X & X & X &  $\cdot$ & $\cdot$  & $\cdot$  & $\cdot$  \\
NS$'$ & X & X & X & X & $\cdot$  & $\cdot$  & $\cdot$  & $\cdot$  & X & X \\
NS$_{\theta}$ & X & X & X & X & \multicolumn{2}{c}{( {X} )$_{\hat{\theta}}$} & $\cdot$  & $\cdot$  &  \multicolumn{2}{c}{( {X} )$_{\theta}$} \\
D6          & X & X & X & X & $\cdot$  & $\cdot$  & $\cdot$  & X  & X & X \\
D6$'$            & X & X & X & X & X & X &  $\cdot$ & X  & $\cdot$  & $\cdot$  \\
D6$_{\theta}$ & X & X & X & X & \multicolumn{2}{c}{( {X} )$_{\hat{\theta}}$} & $\cdot$  & X  &  \multicolumn{2}{c}{( {X} )$_{\theta}$} \\
O6          & X & X & X & X & $\cdot$  & $\cdot$  & $\cdot$  & X  & X & X \\
\end{tabular}
\caption{Summary of the 4d brane setups considered in this paper. We denote by $\hat{\theta}$ the angle $\frac{\pi}{2}-\theta$. When the direction $x_3$ is compactified the 3d brane setup follows from T-duality along this compact direction.}
\label{position}
\end{table}

We have now reviewed the 4d setups and are ready to study the reductions of such dualities on a circle, by compactifying a space direction, say $x_3$.
Following the general discussion of \cite{Amariti:2015yea,Amariti:2015mva,Amariti:2016kat}, we expect that we can provide an effective description of the 4d duality on $S^1$ by employing circle reduction and T-duality along the compact direction.

The brane setup of the reduced theory consists of D3, D5, NS branes and O5 planes.
The fivebrane are wrapped along the compact direction $x_3$ and are extended along the three spacetime directions $x_{012}$ and two extra directions, corresponding to $x_{45}$ for an NS brane and $x_{89}$ for an NS$'$ brane.
Again, NS branes rotated in the planes (4\,5) and (8\,9) correspond to 
NS$_{\pm \theta}$.
The D4 branes stretched between pairs of NS branes become 
D3 after T-duality along $x_3$.
Analogously D6, D6$'$ and D6$_{\pm \theta}$ become 
D5, D5$'$ and D5$_{\pm \theta}$ respectively.

The last ingredient, necessary to engineer the gauge theories in 3d, is understanding the fate of the O6 planes under T-duality.
Following the discussion of \cite{Hanany:2001iy}, an O6 splits into a pair of O5 on $S^1$. We fix the origin of the circle as the point occupied by one of these O5 planes and refer to the other point, symmetric on the circle with respect to the origin, as the mirror point\footnote{The notion of mirror point was originally introduced in \cite{Amariti:2015mva}. In the compact geometry it corresponds to the point symmetric with respect of the origin in $x_3$. The reason why it is denoted as mirror point is not only related to such a symmetry property but also to the fact that in the simple case of $\UU(N)$ SQCD the dual picture can be simplified by locally applying mirror symmetry for the gauge theory associated to the D3 branes at such point.}.
If we also specify the RR charge of the O6 planes, under T-duality we have O6$^+\rightarrow ($O5$^+$,\,O5$^+)$ and 
O6$^-\rightarrow ($O5$^-$,\,O5$^-)$.

In the case at hand, we have an O6$^{+}$ and an O6$^{-}$ separated by an NS$'$ brane, and this setup is consistent only if eight half-D6 branes are added on the O6$^-$ plane.
In the T-dual consistent picture, we then need to split these eight half-D6 branes into four half-D5 at the origin and four half-D5 at the mirror point, placing them on the respective O5$^-$ planes. This way the RR charge conservation is respected when the orientifold crosses the NS$'$ fivebrane.

Before continuing, however, there is a caveat related to the overall $\UU(1)$ gauge symmetry. In the 4d picture discussed above we referred to the gauge group as $\SU(N)$, even though the brane picture provides a $\UU(N)$ gauge group. The reason is dynamical: in the IR the gauge coupling of the $\UU(1)$ factor flows to zero, while the non-abelian part in general  (depending on the number of D6 branes) does not.
In this sense, the $\UU(1)$ gauge freedom corresponds to the freedom of moving the center of mass of the system, and the remnant of this gauge symmetry manifests itself as a baryonic $\UU(1)$ symmetry.
The situation in 3d is different, because the gauge coupling of the abelian factors does not necessarily vanish, and it is natural to interpret the gauge theories as $\UU(N)$ instead of $\SU(N)$.
Such difference is implemented in field theory by gauging the baryonic symmetry.

Once the T-duality rules are specified, we can consider their effect on the brane setup discussed above.
This proved to be an effective description of the model on $\mathbb{R}^{1,2} \times S^1$. 

In the brane picture we have $N$ D3 branes placed at $x_3=0$ and  stretched along $x_6$  between two NS$_{\pm \theta}$ branes, which are wrapped on the compact direction $x_3$.
There is additionally an NS$'$ branes wrapped on $x_3$ and placed on $x_6$ symmetrically with respect to the two NS$_{\pm \theta}$ branes.
At the origin and at the mirror point on the NS$'$ fivebrane we also have the O5$^{+}$/O5$^{-}$ orientifolds, semi-infinite on  $x_7>0$ and $x_7<0$ respectively. Four half-D5 branes are then placed on both the semi-infinite O5$^-$ planes.

Now we  consider the D5, D5$'$ and D5$_{\pm \theta}$. In the first and second case, when  such branes are placed on the orientifold at $x_3=0$, this gives rise to flavour doubling as in the 4d case. In the last case the D5$_{\pm \theta}$ branes  are still placed at $x_3=0$ and in the direction $x_6$ they occupy the same position of the former D6$_{\pm \theta}$ branes.

We can also move some of the D3 and D5 branes along the compact direction, from the origin to the mirror point, splitting $N$ into $N_1+N_2$ and $F$ as $F_1 +F_2$. 

The electric theory is then, in general, a product of two $\UU(M)$ sectors: here $\UU(N_1)$ and $\UU(N_2)$. In each sector there is a conjugate symmetric $\tilde S_{1,2}$ and an antisymmetric $A_{1,2}$ , in addition to four fundamentals  $T_{1,2}$. There are also $F_{1,2}$ fundamental flavors 
denoted as $(Q_{1,2};\tilde Q_{1,2})$.

\begin{figure}
\begin{center}
  \includegraphics[width=14cm]{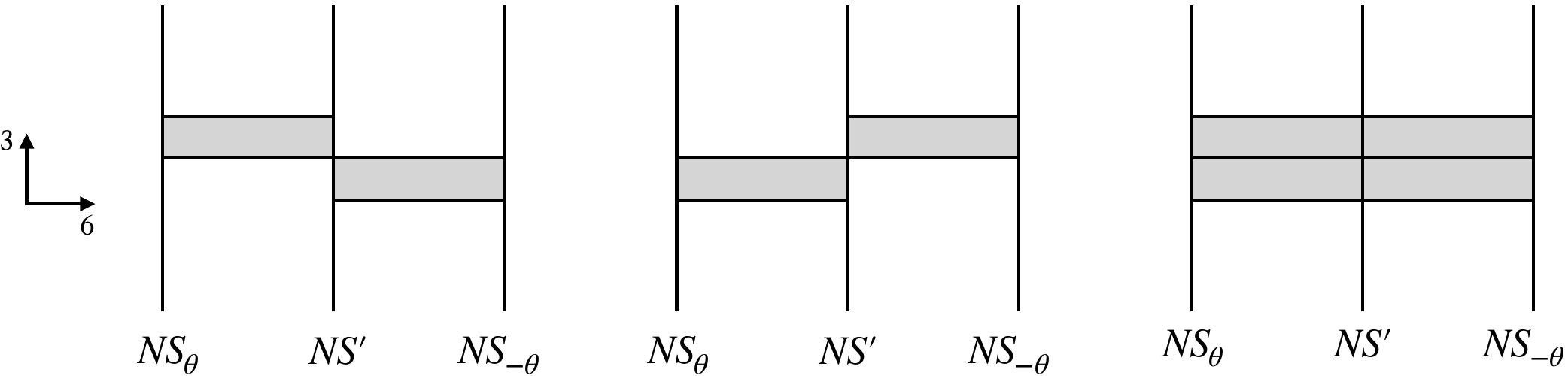}
  \end{center}
  \caption{In these pictures we represent the consistent ways to add Euclidean D1 branes (depicted in grey) to the brane setup in presence of an O5 plane on the central NS$'$ brane. The three configurations are obtained by moving some of  the D3 branes (extended along $x_6$ in the figures) along the direction $x_3$. The first two configurations correspond to monopoles with opposite charges under the topological global symmetry $\UU(1)_J$ arising from the  $\UU(1) \subset \UU(N)$ gauge symmetry. The last configurations is less common and refers to a monopole arising from $\SU(N) \subset \UU(N)$. It is uncharged under the topological symmetry.}
    \label{Fig:monopoles}
\end{figure}

The brane picture suggests that the two sectors interact through an AHW-like superpotential (Affleck-Harvey-Witten) \cite{Affleck:1982as}, due to the euclidean D1 branes stretched between the two stacks of D3 branes and the NS fivebranes along 
$x_3$ and $x_6$. The D1 action gives indeed rise to interactions between the Coulomb branch variables of the two $\UU(M)$ gauge sectors.
Such monopole superpotentials descend from the Higgsing of the original $\UU(N)$ and this is realized in the brane setup by separating the $N$ D3 branes into two stacks, with $N_1$ and $N_2$ D3 respectively.
We give a pictorial representation of the possible configurations in Figure \ref{Fig:monopoles}. 
In the figure we see three types of configurations arising. Two of them are charged (with opposite charge) under the topological symmetry, while the third one is uncharged.

We can build the IR dual description through an HW transition, exchanging the relative positions of the NS$_{\pm \theta}$ branes.
We have different possibilities depending on the type of flavor branes. If we consider D$5_{\pm \theta}$,
we have $3F_1 - N_1+2$ D3 at the origin of the compact coordinate and  $3F_2 - N_2+2$ D3 at the mirror point.
Instead, if we consider either $2F$ D5 or $2F$ D5$'$ branes 
we have $2F_1 - N_1+2$ D3 branes at the origin of the compact coordinate and  $2F_2 - N_2+2$ D3 branes at the mirror point.

The structure of the dual superpotential is identical to the one of the 4d parent dualities in both the gauge sectors. There is additionally an AHW-like superpotential involving the monopoles. It is associated to the Euclidean D1 branes in Figure \ref{Fig:monopoles}.

When the T-dual radius is large, we can effectively focus on the theory at $x_3=0$, decoupling the physics at the mirror point.
Following the prescription of \cite{Amariti:2015mva} we first locally dualize the model at the mirror point, such that it is identical to the one of the electric theory, and then focus on the duality at $x_3=0$. The two sectors at the mirror point are almost identical, except for massless singlets arising from the dual phase. Such singlets are electric monopoles acting as singlets in the dual phase.
Consequently, we arrive to the 3d duality between $\UU(N)$ and $\UU(3F-N+2)$. 
In the following, we will reproduce the results obtained from the brane engineering of the duality from the localization approach of \cite{Amariti:2024bdd} where it was proposed a general recipe to perform the double scaling limit in the reduction of the superconformal index to the partition function.

As anticipated it the beginning of this section, we conclude our analysis  by 
commenting on the cases with an NS brane instead of an NS$'$ brane (See figure \ref{Fig:NSO6}). In such case we can either consider an O6$^+$ or an O6$^-$ plane, which give rise respectively to a conjugate pair of symmetric or antisymmetric two-index tensor fields.
The superpotential interaction is either $W= (S \tilde S)^2$ or $W= (A \tilde A)^2$.
In these cases there are no half-D6 branes on the orientifolds and we can either consider $F$ flavor D6 branes parallel to the orientifold, or D6$'$ parallel to the NS brane or rotate them at an angle in the directions (4\,5) and (8\,9).
The different configurations lead to various  quartic interactions among the fundamentals and the tensors.
Here we focus on the "minimal" configurations, with D6$_{\pm \theta}$, where no further superpotential terms arise.
The other cases are indeed related to this one, because the relative deformations induce a dual Higgsing that is reflected in a different rank for the dual gauge group after the HW transition.

In this case, the dual theory after the HW transition is $\UU(3F-N\pm 4)$ and reproduces the 4d duality. After compactification and T-duality, the expected dual is 
$\UU(3F-N\pm 2)$ with the same monopole structure of the chiral case analyzed above. The analysis for the case with the conjugated antisymmetric tensors has been carried out in full detail in \cite{Amariti:2015mva} and  a similar strategy can be worked out in the symmetric case as well. 

Observe that, in this last case, it is possible to find a configuration on the mirror point of the dual side with a pure $\UU(1)$ gauge symmetry even without rotating any D5 brane at such point on the electric side. This has the same interpretation as a similar situation appearing in the ARSW reduction of SQCD with orthogonal gauge symmetry, where the Coulomb branch on the circle remains unlifted.

\section{Circle reduction and double scaling limit on $Z_{S_b^3}$}
\label{sec:genscaling}

In this section, we study the reduction of the index identity, obtaining an identity between an electric and a magnetic partition function.
We adopt the prescription (and notations) spelled out in \cite{Amariti:2024bdd}, in order to reproduce the behavior found from the D-brane engineering.
Here, we assume the validity of the 4d identity and provide a proof of the validity of the 3d results without any further assumption. 

This is the approach adopted in the derivation of many 3d dualities with a 4d parent. Indeed, one usually reduces the 4d identity on the circle obtaining a 3d identity in presence of a constraint between the 3d real masses. Such constraint corresponds to the balancing condition among the 4d flavor fugacities. While in 4d this constraint is necessary to restrict to the non-anomalous flavor symmetries, in 3d the constraint signals the presence of a monopole superpotential. In the theories obtained from the reduction of the 4d index at vanishing gauge holonomies the corresponding monopole superpotential is indeed the KK monopole superpotential.

For other choices of gauge and flavor holonomies that do not eliminate the 4d constraint, other possible monopole superpotential need to be considered (see e.g. \cite{Benini:2017dud}).
Once a relation between an electric and a magnetic partition function in presence of a balancing condition is obtained, the next step is finding a so called pure 3d duality, corresponding to an identity with unconstrained mass parameters.

At field theory level, the procedure consists of assigning suitable large real masses to some matter fields, and corresponds to the limit of large mass parameters in the integral identities between the partition functions.
In principle, commuting the integral and the limit can require further large shifts for  some of the integration variable.
Such shifts corresponds to Higgs flows on the partition function, and in many cases a large mass flow on one side of the duality requires also an Higgs flow in the dual phase.
Such combined effects give rise to new gauge sectors with possible charged matter 
fields. The instantons of such Higgs flow corresponds to monopoles
interacting through AHW-like superpotentials.
Furthermore, the extra sectors can be dualized in terms of singlets, allowing to reconstruct known pure 3d dualities.

This discussion assumes, at the level of localization, the knowledge of 
the 4d identity and of the 3d identity for the confining duality. 
In general, confining dualities with two-index tensors are not known, and this makes the derivation of the pure 3d identity quite involved (see for example \cite{Amariti:2014iza} for the case of $\UU(N)$ adjoint SQCD for an exception). 
Furthermore, the scheme just described does not necessarily hold when the Coulomb branch is not completely lifted by the KK monopole superpotential. This is for example the case of the reduction of 4d Seiberg-like duality for orthogonal SQCD \cite{Intriligator:1995id} as discussed in \cite{Aharony:2013kma}.  

A different approach consists of considering a double scaling limit of the index for both the gauge and flavor holonomies. We review the basic aspects
of this limit  at the level of the one loop determinants for the chiral and the vector multiplet (also known as  elliptic Gamma functions in the literature) in Appendix \ref{dsrev}.
In this way, by identifying pairs of dual sub-leading saddles, one can work with a compact Coulomb branch and derive well defined identities between product gauge groups in presence of a balancing condition. In these setups, the balancing condition represents a monopole superpotential among the broken gauge sectors.
If some of the dual gauge sectors can be further dualized, one can decouple equivalent gauge sector in the dual phases, ultimately obtaining a pure 3d duality. This approach was employed in \cite{Amariti:2024bdd} to re-derive old results obtained through the ARSW prescription, as well as to find new results, including the orthogonal SQCD example mentioned above.

In this second case, the knowledge of an integral identity for a pure 3d (confining) duality is necessary, and in presence of tensors many dualities of this type are generically not available\footnote{Even if some of them can be proved through tensor deconfinement.} (often they are the limiting cases of the very dualities that one wants to prove in the reduction process). 

\begin{figure}
\begin{center}
  \includegraphics[width=6cm]{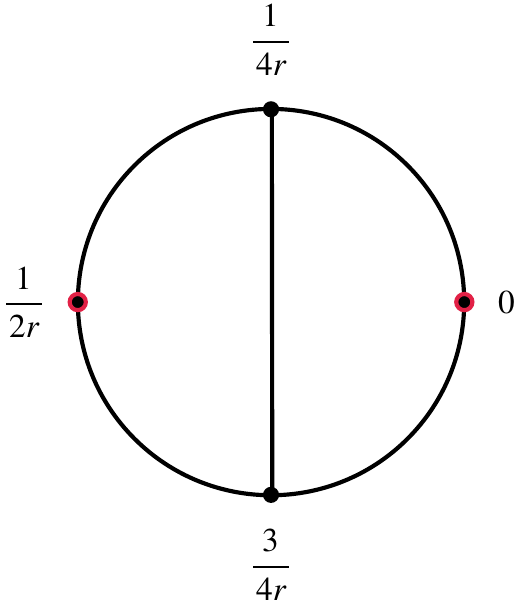}
  \end{center}
  \caption{This is a pictorial representation of the compact direction $x_3$, where $\frac{1}{r}$ represents
  is the periodicity of the compact scalar $\sigma$  on $S^1$.
The red circles at the origin and at the point  $\frac{1}{2r}$, often denoted in the body of the paper as the mirror point, 
represent the positions of the orientifolds in the geometric setup.}
    \label{Fig:Circle}
\end{figure}

When the models admit an HW realization, as the ones discussed above, the brane engineering suggests another strategy for recovering the dualities at the level of the three sphere partition function.
This consists of considering a symmetric configuration with respect to the vertical axis as in Figure \ref{Fig:Circle}.
This requires in the brane picture to have a configuration where the orientifold is either absent or splits into two symmetric orientifolds in the T-dual circle, as in the cases at hand here.

The real masses and the gauge holonomies need to be assigned consistently at the level of the superconformal index in this double scaling approach.
This corresponds to choosing the real masses in the 3d case by pairwise identifying the masses associated with the red circles in Figure \ref{Fig:Circle}, thereby ensuring the validity of the balancing condition.

If we do not fix any flavor and gauge holonomy at $1/4r$ and $3/4r$ mass and gauge, we obtain an identity 
\begin{equation}
\label{squares}
Z_{S^3_b,\text{ele}}^{2} = Z_{S^3_b,\text{mag}}^{2}\, ,
\end{equation}
 which becomes $Z_{S^3_b,\text{ele}} = Z_{S^3_b,\text{mag}}$ in presence of a balancing condition. The balancing condition in such case is not associated in general to a KK monopole but to other types of monopole deformations (like the quadratic monopoles discussed in \cite{Amariti:2018gdc}).


We can also remove the constraints imposed by the monopole superpotential by considering a real mass flow (and, if necessary, a dual Higgs flow) in the $r \rightarrow 0$ limit. Such flow can be engineered in the brane picture by symmetrically moving D5 and possibly D3 branes on the vertical axis in Figure  \ref{Fig:Circle}.
In this way we maintain the structure of the identity (\ref{squares}) also after the real flow is realized. At field theory level indeed we need to choose a suitable assignment of the real masses, compatible with the balancing condition and the broken symmetry structure,  that does indeed keep an identity of type (\ref{squares}).

Such an identity is obtained then by considering  the two gauge sectors on the vertical axes in (\ref{squares}). Observe that they  do not have any symmetric or antisymmetric two-index tensor, because  of the absence of orientifolds in the brane picture at such positions on the circle. Indeed the former tensors become bifundamentals connecting the two gauge groups,
one placed at $x_3 = \frac{1}{4r}$ and the other at  $x_3 = \frac{3}{4r}$. 
This gives rise to a quiver with two unitary gauge factors connected by a pair of conjugated bifundamentals.
Such quiver can then be locally dualized, using elementary Aharony dualities \cite{Aharony:1997gp}
 for unitary SQCD, and the resulting configuration yields an identity of the form  (\ref{squares}), where on the dual side the "typical" contribution of electric monopoles acting as singlets in the dual phase appears explicitly in the partition function.
Along these lines in the following we present the explicit realization of the duality from the double scaling, obtaining the quadratic  relation of the  type \eqref{squares} and interpreting it at physical level.

\subsection{Example I: $\UU(N)$ with $A$ and $\tilde A$}

We start our analysis with the duality involving a $\UU(N)$ gauge group with a conjugated pair of antisymmetric tensors interacting through a quartic superpotential and $F$ pairs of fundamental and antifundamental flavors.

From the field theory perspective the duality can be reduced to 3d by exploiting the ARSW prescription and by considering $F+1$ fundamental flavors. One can then assign  large real masses of opposite sign to a fundamental and to an antifundamental on the electric side. On the dual side the corresponding vacuum is realized by performing a further Higgs flow assigning a large vev to a scalar in the vector multiplet, breaking the $\UU(3F-N-1)$ gauge symmetry to $\UU(3F-N-2) \times \UU(1)$.

The $\UU(3F-N-2)$ gauge sector has two conjugated antisymmetrics interacting through a quartic superpotential in addition to  $F$ pairs of dual fundamental flavors and the same mesonic flippers of the 4d model  (i.e. the 3d superpotential for this sector is identical to the 4d one given in (\ref{Wdualaat})). In the $\UU(1)$ sector there are a fundamental and an antifundamental interacting with a singlet that remains massless, that arises from the electric singlet $Q_{F+1} A \tilde A \tilde Q_{F+1}$.
Observe that there is a second massless singlet $Q_{F+1} \tilde Q_{F+1}$ that is not interacting with any charged  matter field.
This singlet interacts with a monopole, that we denote as $v_0$, arising from the \mbox{$\UU(3F-N-2)$} sector, associated to the Coulomb branch $(1,0,\dots,0,-1)$.
In addition the real mass flow removes the KK monopoles and it adds an AHW interaction between the \mbox{$\UU(3F-N-2)$} and the $\UU(1)$ gauge sectors.
We can further dualize the $\UU(1)$ SQED sector. The meson is massified by the superpotential interaction with the singlet arising from $Q_{F+1} A \tilde A \tilde Q_{F+1}$ while the two monopoles and anti-monopoles charged under $\UU(1)_J$ give rise to two interactions flipping the flux $+1$ monopole $v_+$  and the flux $-1$ anti-monopole $v_-$   of the $\UU(3F-N-2)$ sector.

The final duality can be summarized as follows
\begin{equation}
\label{AAtpure3d}
\begin{array}{|cc|}
\hline
\UU(N) & \UU(3F-N-2) \\
(A,\tilde A) \oplus F(Q,\tilde Q)& (a,\tilde a) \oplus  F (q,\tilde q) \\
W = (A \tilde A)^2 & W = (a \tilde a)^2 + P \tilde q^2 a+ \tilde P q^2 \tilde a \\
& \phantom{petosino} + M_0 q a \tilde a \tilde q + M_1 q \tilde q + v_{\pm,0} V_{\mp,0}\\ 
\hline
\end{array}
\end{equation}
where the last term in the dual $W$ is a shortcut for $v_{+} V_{-} +
v_{-} V_{+} +  v_{0} V_{0}$ (we will adopt this conventions in the other examples below as well).
The mesonic singlets of the dual phase correspond to $P = \tilde A  Q^2$,
$\tilde P = A  \tilde  Q^2$, $M_0 = Q \tilde Q$ and $M_1 = Q  A \tilde A \tilde Q$.
The singlets $V_{\pm}$ and $V_0$ can be identified with the monopoles of the electric phase acting as singlets in the dual phase, flipping the relative dual monopole operators $v_{\pm,0}$.

Following the ARSW prescription at the level of the three sphere partition function,
we have the identity
\begin{eqnarray}
\label{ARSWonAAt}
&&
Z_{\UU(N)}^{(F\square; F \overline \square; 1 \overline{A};1 A)}(\vec \mu;\vec \nu; \tau_{\tilde A};\tau_A;\Lambda) 
=
\prod_{1\leq j< \ell \leq F}
\Gamma_h(\tau_{\tilde A} +\mu_j +\mu_\ell,\tau_A+\nu_j +\nu_\ell)
\nonumber \\
&& \times
\prod_{1\leq b, c \leq F}
\Gamma_h(\mu_b+ \nu_c; \mu_b+ \nu_c +\tau_A +\tau_{\tilde A})\;
Z_{\UU(3F-N-4)}^{(F \square; F \overline \square; 1 \overline{A} ;1 A)}\left(\vec {\tilde \mu};\vec {\tilde \nu}; \tilde  \tau_{\tilde a};\tilde  \tau_a;-\Lambda\right),
\end{eqnarray}
which holds if we impose the constraint
\begin{equation}
\label{bcaat}
\sum_{b=1}^{F} (\mu_b+\nu_b)=(2F-N-2)\omega\,.
\end{equation}
Furthermore the mass parameters $\tau_A$ and $ \tau_{\tilde A}$ are
\begin{equation}
\tau_A = x+\frac{\omega}{2},\quad \tau_{\tilde A }= -x+\frac{\omega}{2}\,.
\end{equation}
 The arguments appearing in the dual  partition function $Z_{\UU(\tilde N)}$ are related to the ones on the electric side by the dictionary 
\begin{eqnarray}
\label{dictionary}
\tilde \mu_{b} &=& \frac{\omega}{2} -\mu_b + \frac{F-2}{\tilde N} x \,, \nonumber \\
\tilde \nu_{b} &=& \frac{\omega}{2} -\nu_b - \frac{F-2}{\tilde N} x\,, \nonumber \\
\tilde \tau_a &=& \frac{\omega}{2} + \frac{N-F}{\tilde N} x\, , \nonumber \\
\tilde \tau_{\tilde a} &=& \frac{\omega}{2} - \frac{N-F}{\tilde N} x\,.
\end{eqnarray}

We can then remove the constraint by considering $F+1$ flavors and assigning  large real masses with opposite sign to a fundamental and to an antifundamental, as discussed above.
In the dual, we are left with a $\UU(1)$ SQED sector that can be dualized in terms of three singlets. After substituting the constraint \eqref{bcaat}, we arrive at the final identity 
\begin{eqnarray}
\label{antisymmetricidons1}
&&
Z_{\UU(N)}^{(F\square; F \overline \square;1  \overline{A};1 A)}(\vec \mu;\vec \nu; \tau_{\tilde A};\tau_A;\Lambda) \;
=
\prod_{1\leq b< c \leq F}
\Gamma_h(\tau_{\tilde A} +\mu_b +\mu_c,\tau_A+\nu_b +\nu_c)
\nonumber \\
&& \times
\prod_{1\leq b,c \leq F}
\Gamma_h(\mu_b+ \nu_c; \mu_b+ \nu_c +\tau_A +\tau_{\tilde A})\;
\Gamma_h\left((2F-N)\omega-\sum_{b=1}^{F} (\mu_b+\nu_b)\right)
 \\
&& \times\;
\Gamma_h\left(\pm \frac{\Lambda}{2} +\frac{2F-N+1}{2}  \omega-\frac{1}{2}\sum_{b=1}^{F} (\mu_b+\nu_b)\right)
Z_{\UU(3F-N-2)}^{(F \square; F \overline \square; 1 \overline{A} ;1 A)}\left(\vec {\tilde \mu};\vec {\tilde \nu}; \tilde  \tau_{\tilde a};\tilde  \tau_a;-\Lambda\right)\,, \nonumber
\end{eqnarray}
which holds in absence of any further constraint.
The duality dictionary given in (\ref{dictionary}) can still be used in this identity, with the proper identifications of $N$, $F$ and $\tilde N$ with the ones of this phase.

A comment is in order. The normalization of the  $\UU(1)_X$ is ambiguous because we can always shift the gauge symmetry by a finite amount and re-absorb such a shift in the $\UU(1)_X$  symmetry. The net effect of the shifts consists then of an overall pure phase factor that does not spoil the physical identity between $|Z_{ele}|$ and $|Z_{mag}|$.
Having this ambiguity in mind in the rest of the paper we will always normalize the $\UU(1)_X$ charge as in 4d, which in the $\SU(N)$ case follows from the correct identification of baryons under the duality.
The same duality derived from the ARSW prescription can be derived in an alternative way by using a double scaling limit. This limit will be quite useful when we will analyze the other examples in the next sections and for this reason we study it here, demonstrating the exact matching with the results obtained from the ARSW prescription.
We will discuss two possible ways of deriving the duality from the double scaling limit.
\begin{itemize}
\item \underline{Double scaling (I)}\\ \\
The first case is conventional and corresponds to the same one discussed in \cite{Amariti:2024bdd}.
It consists of studying an effective model on $S^1$ by turning on some holonomies for the flavor and gauge symmetries.
We start with $F+2$ flavors and assign two opposite real masses to two pairs of them, without Higgsing the gauge group. The explicit values of such masses scale with the inverse radius of the circle and correspond to considering the configurations with shifts $\pm \frac{1}{2r}$. On the other hand, in the dual side we also Higgs the gauge group, breaking it into $\UU(3F-N-2)\times \UU(4)$.
The $\UU(4)$ group is obtained by shifting four real masses at $\frac{1}{2r}$, where we also have two massless conjugated antisymmetric fields in the spectrum in addition to the two flavors and to the leftovers of the flippers $P,\tilde P,M_0$ and $M_1$.
The $\UU(4)$ sector has the same superpotential of the model at $\sigma=0$ and furthermore the two gauge sector interact through an AHW interaction of the type discussed above. 
As discussed in  Appendix \ref{app:U4} this sector is confining and, turning to the dual description for this sector, we obtain the same duality obtained from ARSW.
At the level of the three sphere partition function we obtain again the identity \eqref{antisymmetricidons1}.

\item \underline{Double scaling (II)}\\ \\
 As discussed in Section \ref{sec:genscaling}, there is also another way to work out the duality from the double scaling limit of the superconformal index. This approach is less physical but it leads to the same identity at the level of the partition function.
 It consists of finding a "symmetric" configuration of flavor and gauge fugacities
 with respect to the vertical axis in Figure \ref{Fig:Circle}.
 The minimal case\footnote{Actually we could also consider the flavor holonomies only at $0$ and $1/2r$ but such a procedure gives rise to an identity with a balancing condition between mass parameters, signalling the presence of a (non -KK) monopole superpotential.  In this case, with an antisymmetric flavor, this identity is well defined and, removing the monopole superpotential by real mass flows, gives rise to the pure 3d duality found above. Similar constructions apply to the cases below, but they do not necessarily give rise to non-divergent partition functions. We will comment on such possibility in the next sections.} consists of choosing $\UU(2N)$ with $2F+2$ fundamentals on the electric side and displace the $F$ flavor fugacities at the origin and at $1/2r$ and the other two at $1/4r$ and $3/4r$. The gauge group in this case is split into two $\UU(N)$ factors at the origin and at $1/2r$. The dual gauge group is then broken into two 
 $\UU(3F-N-2)$ gauge factors at the origin and at $1/2r$ and two $\UU(3)$ factors at $1/4r$ and $3/4r$.  

Each $\UU(3)$ factor has a fundamental flavor and interacts with the other $\UU(3)$ through two  conjugated bifundamentals, arising from the two antisymmetric tensors.  
These fields also inherit a quartic interaction from the original $(A \tilde A)^2$ superpotential term.
The presence of the two orientifolds  in the brane description is reflected in
the fact that the two FI of the two $\UU(3)$ factors are~equal.

The model also includes a further superpotential involving flippers, which interact with the charged matter fields. 
Explicitly, the $\UU(3)\times \UU(3)$ model is described by the quiver in Figure \ref{Fig:quivU3} and the confining duality is studied in the appendix.
Referring to that analysis  we conclude that the $\UU(3) \times \UU(3)$ sector can be dualized in six singlets, corresponding to the contributions of the three monopoles $V_{\pm,0}$, which in this symmetric situation appear twice, i.e. they interact with both the $\UU(3F-N-2)$ sectors.

Correspondingly, we arrive to the following identity between partition functions
\begin{eqnarray}
\label{quasiquadratoperfetto}
&&
\left(
Z_{\UU(N)}^{(F\square; F \overline \square; 1 \overline{A};1 A)}(\vec \mu;\vec \nu; \tau_{\tilde A};\tau_A;\Lambda) \right)^2
=
\prod_{1\leq b<c \leq F}
\Gamma_h^2(\tau_{\tilde A} +\mu_b +\mu_c,\tau_A +\nu_b +\nu_c)
\nonumber \\
&& \times
\prod_{1\leq b,c \leq F}
\Gamma_h^2(\mu_b+ \nu_c; \mu_b+ \nu_c +\tau_A +\tau_{\tilde A})
\left(Z_{\UU(3F-N-2)}^{(F \square; F \overline \square; 1 \overline{A} ;1 A)}
\left(\vec {\tilde \mu};\vec {\tilde \nu}; \tilde  \tau_{\tilde A};\tilde  \tau_A;-\Lambda
\right)\right)^2
\nonumber \\
&& \times \;
\Gamma_h(\tau_{\tilde A}\!+\!\mu_{F+1}\!+\!\mu_{F+2} )\,
\Gamma_h(\tau_A\!+\!\nu_{F+1}\!+\!\nu_{F+2})\!\!
\prod_{b=F+1}^{F+2}\!\! \Gamma_h(\mu_b\!+\! \nu_b; \mu_b\!+\! \nu_b\!+\!\tau_A \!+\!\tau_{\tilde A})
\nonumber \\
&& \times
\int \prod_{i=1}^3  \ee^{-\mi \pi \Lambda (\rho_i+\eta_i)}  \dd{\rho_i} \dd{\eta_i}
\Gamma_h( \tilde \mu_{F+1}-\rho_i ,\tilde \nu_{F+1} +\rho_i )\,
\Gamma_h( \tilde \mu_{F+2}-\eta_i ,\tilde \nu_{F+2}+ \eta_i )   
\nonumber \\
&& \times \;
\frac{\prod_{i,\ell=1}^3 \Gamma_h(\tilde \tau_A +\rho_{\ell}+\eta_{i},
\tilde \tau_{\tilde A}-\rho_{\ell}-\eta_{i})}
{\prod_{1\leq i<\ell \leq 3}
\Gamma_h(\pm(\rho_{i}-\rho_{\ell}),\pm(\eta_{i}-\eta_{\ell}))}\,,
\end{eqnarray}
 where the last three lines correspond to the $\UU(3) \times \UU(3)$ sector. 
 
  Again the duality dictionary follows from (\ref{dictionary}), with the proper identifications of $N$, $F$ and $\tilde N$ with the ones of this phase, and in addition here the balancing condition reads
\begin{equation}
\label{BCsquare}
 \sum_{b=1}^{F} (\mu_b+\nu_b) + \frac{1}{2}(\mu_{F+1} + \mu_{F+2} + \nu_{F+1} + \nu_{F+2} )=  \omega(2F-N+1)\,.
\end{equation}
The $\UU(3)\times \UU(3)$ sector can be studied along the lines of the discussion in the Appendix \ref{app:0} and, upon fixing $\Lambda_1=\Lambda_2$ and $\mu_{F+1} + \nu_{F+1} = \mu_{F+2} + \nu_{F+2} $, can be traded with the monopoles in formula (\ref{finalU3U3}).
The net result is that the last three lines of \eqref{quasiquadratoperfetto} become
\begin{equation}
\label{finalU3U3text}
\Gamma_h^2\left(-(\tilde \mu_{F+1}+\tilde \nu_{F+1})\right)
\Gamma_h^2\left(
\pm \frac{\Lambda}{2} +\frac{\omega}{2}-\frac{1}{2} (\tilde \mu_{F+1}+\tilde \nu_{F+1}) \right).
\end{equation}
Plugging this result in \eqref{quasiquadratoperfetto},
we observe that the identity is of the form 
\eqref{squares} and after extracting the square roots we arrive again at \eqref{antisymmetricidons1}, where the three monopoles $V_{\pm,0}$ arise from  \eqref{finalU3U3text} after applying the balancing condition \eqref{BCsquare}.

In the next sections, we will use this last prescription to study the 4d/3d reduction of the dualities with the conjugated symmetric pair and with the conjugated symmetric and the antisymmetric. This is because, in both cases, the ARSW prescription cannot be adapted to the analysis of the partition function, and a double scaling limit is necessary. Nevertheless, we have not found confining dualities useful to dualize the dual sectors at the  mirror point in presence of one or two symmetric tensors, or in the brane language in presence of O6$^+$ planes. 

\end{itemize}

\subsection{Example II: $\UU(N)$ with $S$ and $\tilde S$}

Here, we study the reduction using the double scaling limit in two different ways, and we show again that the pure 3d limits obtained in both cases agree.
In the first case, we consider $F$ fundamentals and, on the electric side, we keep the flavor symmetry unbroken. 
We also consider a $\UU(N+1)$ gauge theory and the vacuum we analyze consists of an unbroken $\UU(N)$ at the origin and a $\UU(1)$ gauge theory at the mirror point.
The $\UU(1)$ sector has two fields with charge $\pm 2$, corresponding to the original symmetric tensors, shares the same FI term as the $\UU(N)$ sector and contains a quartic superpotential between the two fields.
Similarly the 4d dual model corresponding to a $\UU(3F-N+3)$ gauge theory splits into $\UU(3F-N+2)\times \UU(1)$ on the circle and the $\UU(1)$ sector is identical to that of the electric theory.
There is only a different normalization for the $\UU(1)_X$ symmetry that can nevertheless be absorbed in a finite shift of the real scalar in the vector multiplet.
It follows that the two $\UU(1)$ sectors can be simultaneously removed and one is left with a duality between $\UU(N)$ and $\UU(3F-N+2)$, with a further constraint on the masses of the fundamentals that signals the presence of a monopole superpotential 
(which differs from the KK monopole).
The identity we arrive to at the level of the three sphere partition function is
\begin{eqnarray}
\label{symmetricidons1}
&&
Z_{\UU(N)}^{(F\square; F \overline \square; 1\overline{S};1S)}(\vec \mu;\vec \nu; \tau_{\tilde S};\tau_S;\Lambda) 
=
\prod_{1\leq b\leq c \leq F}
\Gamma_h(\tau_{\tilde S} +\mu_b +\mu_c,\tau_S+\nu_b +\nu_c)
\nonumber \\
&& \times
\prod_{1\leq b,c \leq F}
\Gamma_h(\mu_b+ \nu_c; \mu_b+ \nu_c +\tau_S +\tau_{\tilde S})
Z_{\UU(3F-N+2)}^{(F \square; F \overline \square; 1\overline{S} ;1S)}\left(\vec {\tilde \mu};\vec {\tilde \nu}; \tilde  \tau_{\tilde s};\tilde  \tau_s;-\Lambda\right),
\end{eqnarray}
with the further constraint
\begin{equation}
\label{balancingsst}
\sum_{b=1}^{F} (\mu_b+\nu_b) = (2F-N+1) \omega\,.
\end{equation}
This corresponds to a duality with the same field content and superpotential of the 4d theory and in addition a superpotential $W \propto V_+ V_-$ (defined below) on the electric  side and $W \propto v_+ v_- $ on the magnetic side.

The pure 3d limit can be obtained by removing such monopole superpotentials, integrating out two fundamentals and two antifundamentals with opposite real masses for each representation.
For the ease of notation, this flow can be engineered by considering $F+2$ pairs of fundamentals and antifundamentals, such that the electric theory after the flow is again $\UU(N)$ with F fundamental flavors and a symmetric flavor with the quartic superpotential. On the other hand, the vacuum of the dual theory, that preserves the duality, breaks the gauge group as $\UU(3F-N+2) \times \UU(3)^2$ and the $\UU(3)$ sectors interact through a pair of conjugated bifundamentals, corresponding the the former pair of symmetric and conjugated symmetric tensors. 
The analysis of this sector is similar to the one performed in the section above, but in this case we have two opposite shifts in the FI in the two $\UU(3)$ gauge group. 
The FI are shifted indeed by the quantity $\frac{1}{2} (\mu_{F+1} + \mu_{F+2} +\nu_{F+1} + \nu_{F+2})-\omega$
where the masses are constrained by the balancing condition
\begin{equation}
\sum_{b=1}^{F+2} (\mu_b +\nu_b) = (2F-N+5)\omega\,.
\end{equation}
If we further identify $\mu_{F+1} =  \mu_{F+2}$ and $\nu_{F+1} =  \nu_{F+2}$  and dualize the $\UU(3)\times \UU(3)$ sector using 
\eqref{finalU3U3}, we observe that three monopoles are massive (one has mass parameter proportional to $\omega$ and the other two have parameters $\omega\pm\frac{\Lambda}{2}$) and they evaluate  to $1$ in the partition function because of the inversion relation $\Gamma_h(2\omega-x) \Gamma_h(x)=1$. 
The final identity for the three sphere partition function is 
\begin{eqnarray}
\label{symmetricidons2}
&&
Z_{\UU(N)}^{(F\square; F \overline \square; 1 \overline{S};1 S)}(\vec \mu;\vec \nu; \tau_{\tilde S};\tau_S;\Lambda) 
=
\prod_{1\leq b\leq c\leq F}
\Gamma_h(\tau_{\tilde S} +\mu_b +\mu_c,\tau_S+\nu_b +\nu_c)
\nonumber \\
&&
\times
\prod_{1\leq b,c \leq F}
\Gamma_h(\mu_b+ \nu_c; \mu_b+ \nu_c +\tau_S +\tau_{\tilde S})\,
\Gamma_h\left((2F-N+2)\omega-\sum_{b=1}^{F} (\mu_b+\nu_b)\right)
\nonumber \\
&& \times\,
\Gamma_h\left(\pm \frac{\Lambda}{2} +\frac{2F-N+3}{2}  \omega-\frac{1}{2}\sum_{b=1}^{F} (\mu_b+\nu_b)\right)
Z_{\UU(3F-N+2)}^{(F \square; F \overline \square; 1 \overline{S} ;1 S)}
\left(\vec {\tilde \mu};\vec {\tilde \nu}; \tilde  \tau_{\tilde s};\tilde  \tau_s;-\Lambda
\right),
\nonumber \\
\end{eqnarray}
without further constraint on the mass parameters. Observe that the duality map on the real masses follows again from the 4d one and we keep the same normalization on the mass parameter associated to the $\UU(1)_X$ global symmetry in the dual side.
The final duality can be summarized as follows
\begin{equation}
\label{SStpure3d}
\begin{array}{|cc|}
\hline 
\UU(N) & \UU(3F-N+2) \\
(S,\tilde S) \oplus F(Q,\tilde Q)& (s,\tilde s) \oplus  F (q,\tilde q) \\
W = (S \tilde S)^2 & W = (a \tilde a)^2 + P \tilde q^2 s+ \tilde P q^2 \tilde s \\
& 
\phantom{petosino} + M_0 q s \tilde s \tilde q + M_1 q \tilde q + v_{\pm,0} V_{\mp,0}\\ 
\hline
\end{array}
\end{equation}
where the mesonic singlets of the dual phase correspond to $P = \tilde S  Q^2$,
$\tilde P = S  \tilde  Q^2$, $M_0 = Q \tilde Q$ and $M_1 = Q  S\tilde S \tilde Q$.
Furthermore, the singlets $V_{\pm}$ and $V_0$ correspond to the monopole of the electric phase acting as singlets in the dual phase flipping the relative dual monopole operators $v_{\pm,0}$.

This identity is also useful to explain the failure of the ARSW prescription for the reduction of the index on the three sphere partition function in this case.
If we plug onto the electric side the real masses, that arise from the 4d fugacities constrained as in \eqref{BCSSt4d}, we have the constraint
\begin{equation}
\label{BCSSt3d}
\sum_{b=1}^F (\mu_b+\nu_b) = (2F-N+2)\omega\,.
\end{equation}
Plugging this value onto the magnetic side we observe the emergence of a divergence in the hyperbolic Gamma function, due to the $V_0$ monopole.
This is the same situation that arises in the 4d/3d reduction of the duality for orthogonal SQCD. In the brane picture, the divergence originates from the third D1 brane in Figure \ref{Fig:monopoles}, which is indeed the same setup that one would have obtained removing the central NS brane, i.e. considering $\SO(N)$ SQCD.

We conclude our discussion by applying the double scaling limit and reducing to the pure 3d case at the level of the partition function, thereby obtaining an identity of the form \eqref{squares}.
The discussion is very similar to the one performed in the section above, and for this reason our presentation here will be sketchy.
We start by considering a $\UU(2N)$ gauge theory with $2F+2$ flavors and by placing the $F$ flavor fugacities at the origin and at $1/2r$ and the other two at $1/4r$ and $3/4r$. 
The gauge group in this case splits into two $\UU(N)$ factors at the origin and at $1/2r$. The dual gauge group is then broken into two 
 $\UU(3F-N+2)$ gauge factors, one at the origin and the other at $1/2r$ and two $\UU(3)$ factors at $1/4r$ and $3/4r$.  

Again, the divergencies are removed, and we can fix the real masses symmetrically, obtaining an identity constrained by a balancing condition. Dualizing the $\UU(3)\times \UU(3)$ quiver we obtain a relation of the type \eqref{squares} that becomes exactly \eqref{symmetricidons2} after removing the squares.

\subsection{Example III: $\UU(N)$ with $A$ and $\tilde S$}

The last example concerns the model with a conjugate symmetric and an antisymmetric. In this case, the brane picture plays a crucial role in the search for a  configuration that preserves the duality. This is due to the presence of the half-D6 branes on the orientifold, which force us to consider, in the T-dual picture, a split of these eight fundamentals into two pairs of four fundamentals, at the origin and at the mirror point.

This configuration can be obtained at the level of the double scaling through an opportune real mass flow, that can be engineered at the level of field theory using the $\SO(8) \rightarrow SO(4) \times \SO(4)$ symmetry breaking pattern.

It implies that, even in the simplest cases, where we do not assign any real mass to the other $F$ flavors we have to consider at $x=1/2r$ such four fundamentals interacting with a conjugate symmetric tensor (even if the gauge group is broken to $\UU(1)$ such a field remains in the low energy spectrum as a field with charge $-2$).

The simplest case on the electric side consists of considering a $\UU(N)$ gauge theory with $F+4$ fundamentals and $F$ antifundamentals in addition to the tensors at $x=0$. On the dual side, there is a $\UU(3F-N-2)$ gauge theory at $x=0$ and an extra gauge sector at $x=1/2r$, corresponding to $\UU(2)$ with a conjugated symmetric, an antisymmetric (which has dimension one and survives in the $\UU(2)$ case, having non trivial character), together with four fundamentals interacting with the conjugated symmetric.

We expect this model to be confining and that, by removing it, the expected 3d result is obtained. We have not found a direct proof of this fact (by gauging the baryonic symmetry of other confining dualities or from tensor deconfinement), but we can observe that it is the limiting case of the pure 3d duality that we are looking for. 
We will see, that, \emph{a posteriori}, i.e. by using the identity that we are going to derive using another method, we can dualize the $\UU(2)$ sector and obtain the expected result.
Similar situations arise if we consider some fundamental flavor at $x=1/2r$ or if we consider other vacua on the electric side.
It is in principle  possible to consider on the electric side a $\UU(N+1)$ gauge theory and decouple a $\UU(1)$ sector in both the electric and the magnetic side, obtaining an effective duality with a balancing condition imposed by a monopole superpotential. Nevertheless, such relations are divergent and cannot be used to flow to pure 3d dualities at the level of the partition function. Once again, the divergence can be understood explicitly \emph{a posteriori} from the final formula of the pure 3d duality obtained below.

On the other hand the third approach used in the examples above, consisting of finding a relation of the type (\ref{squares}), works in this case as well. 
We start by considering a $\UU(2N)$ gauge theory with $2F+2$ flavors in addition to the eight fundamentals interacting with the conjugated symmetric. We then displace the $F$ flavor fugacities at the origin and at $1/2r$ and the other two at $1/4r$ and $3/4r$. 
The eight flavors are split symmetrically at the origin and at the mirror point.
The gauge group in this case splits into two $\UU(N)$ factors at the origin and at $1/2r$. The dual gauge group is then broken into two 
 $\UU(3F-N+2)$ gauge factors at the origin and at $1/2r$ and two $\UU(3)$ factors at $1/4r$ and $3/4r$.  
 
 The final relation for the partition functions reads
 \begin{eqnarray}
\label{final}
&&
Z_{\UU(N)}^{[(F+4)\square; F \overline \square;  1 \overline{S};1  A]}(\vec \mu,\vec \zeta;\vec \nu; \tau_{\tilde S};\tau_A;\Lambda)
=
\Gamma_h(\omega(2F-N+2)-\sum_{b=1}^{F} (\mu_b+\nu_b)) 
 \nonumber \\
&& \times \,
\Gamma_h \left(\pm \frac{\lambda}{2}
+ \frac{\omega(2F-N+3)}{2}-\frac{1}{2}\sum_{b=1}^{F} (\mu_b+\nu_b) \right)
\prod_{1 \leq b \leq c \leq F} \Gamma_h(\tau_{\tilde S} +\mu_b +\mu_c)
\nonumber \\
&& \times
\prod_{1 \leq b < c \leq F}
\Gamma_h(\tau_A +\nu_b +\nu_c)
\prod_{b,c=1}^{F} \Gamma_h(\mu_b+ \nu_c; \mu_b+ \nu_c +\tau_A +\tau_{\tilde S})
\nonumber \\
&& \times \,
\prod_{b=1}^{F}\prod_{\ell=1}^4 \Gamma_h(\nu_b+ \zeta_\ell) \;
Z_{\UU(3F-N+2)}^{[(F+4)\square; 	F \overline \square; 1 \overline{S};1A ]}
\left(
  \vec {\tilde \mu},\vec {\tilde \zeta};\vec {\tilde \nu}; \tilde  \tau_{\tilde S};\tilde  \tau_A;-\Lambda
\right).
\end{eqnarray}
The final duality can be summarized as follows
\begin{equation}
\label{chiralpure3d}
\begin{array}{|cc|}
\hline
\UU(N) & \UU(3F-N+2) \\
(A,\tilde S) \oplus F(Q,\tilde Q)\oplus4 T& (a,\tilde s) \oplus  F (q,\tilde q)\oplus4t \\
W = (A \tilde S)^2 + \tilde S T^2 & W = (a \tilde s)^2 +\tilde s t^2+ P \tilde q^2 a+ \tilde P q^2 \tilde s \\
& 
\phantom{petosinooo}+ M_0 q a \tilde s \tilde q + N q \tilde q +L t a\tilde s  \tilde q+ v_{\pm,0} V_{\mp,0}\\ 
\hline
\end{array}
\end{equation}
where the mesonic singlets of the dual phase correspond to $P = \tilde S  Q^2$,
$\tilde P = A  \tilde  Q^2$, $M = Q \tilde Q$, $N = Q  S\tilde S \tilde Q$ and $L=T \tilde Q$.

Observe that the  masses for the four extra fundamentals in the partition function  
are parametrized as $\zeta_{1} = \xi_1+\frac{3}{4} \omega$,
$\zeta_{2} = -\xi_1-\frac{3}{4} \omega$,
$\zeta_{3} = \xi_2+\frac{3}{4} \omega$ and $\zeta_{4} = -\xi_2+\frac{3}{4} \omega$.
Furthermore the singlets $V_{\pm}$ and $V_0$ correspond to the monopole of the electric phase acting as singlets in the dual phase flipping the relative dual monopole operators $v_{\pm,0}$.

\section{Conclusions}
\label{sec:end}

In this paper, we have studied the 4d/3d reduction of dualities with a $\UU(N)$ gauge group, two rank-two tensors, fundamentals and fundamentals, using a field theoretical approach, T-duality on the brane setup and  performing an opportune limit on the supersymmetric index. 
We can summarize our main findings in 3d by referring to the three dualities 
\eqref{AAtpure3d}, \eqref{SStpure3d} and  \eqref{chiralpure3d}.
In the analysis we also found other dualities with monopole superpotential turned on, that are crucial for obtaining the correct 3d limit through real mass and Higgs flows that preserve the duality.

Various extensions and further analysis are possible.
A first check that we did not perform here consists of matching at low ranks the expansions of the 3d superconformal index defined in \cite{Kim:2009wb,Imamura:2011su}.
Further checks the dualities found here regard the consistency of various limits and  RG flows.
In addition the confining limits of the new dualities need, in general, a separated analysis, where non perturbative potentials can be generated, similarly to the four dimensional case \cite{Seiberg:1994bz,Csaki:1996zb,Csaki:1998fm,Klein:1998uc,Klein:2003wa}.
An example of such limiting case corresponds to the $\UU(4)$ case studied in Appendix \ref{app:U4}.
In the case at hand, we obtained this confining duality by gauging a baryonic symmetry and adding a superpotential deformation from a confining duality for the $\SU(4)$ case. Beyond the cases analyzed here, similar construction are possible starting from 
 other $\SU(N)$ confining dualities with tensors, as the ones studied in \cite{Csaki:2014cwa,Amariti:2015kha,Nii:2019ebv, Benvenuti:2021nwt, Amariti:2022wae,Amariti:2024gco}.
Furthermore, it may be useful to derive such cases independently from tensor deconfinement.

Another generalization of the analysis performed here consists in studying the models with the addition of an adjoint tensor $\Phi$, with superpotential 
$\Phi^{k+1}$. A brane engineering indeed exists for generic $k$, it is obtained by increasing the number of fivebranes.
It would be interesting to generalize our analysis to such case, where a  more monopole structure is expected, due to the possible dressing with the adjoint.
Another brane setup that could be interesting to include in our analysis consists  of considering $\tilde O5^{\pm}$ planes instead of $O5^{\pm}$. Similar setups have been recently studied in the literature in a different framework  (see \cite{Huertas:2024mvy}). They require the presence of additional half-branes and possibly CS terms and they can give origin in 3d to other types of dualities, different from the ones derived here.

Moreover, one may also study the reduction of other 4d dualities without a brane engineering, as the case with a conjugate symmetric and an antisymmetric without the superpotential deformation $\tilde S T^2$. Our prescription should still apply, since the deformation can be removed in 4d by adding a mass term of the type $T \tilde Q$, and an analogous flow can be implemented directly in the pure 3d duality studied here.
From the pure 3d theories found here one may also gauge the topological symmetry, obtaining $\SU(N)$ dualities. In the dual case $\UU(\tilde N) \times \UU(1)$
gauge theories are expected, where the $\UU(1)$ sectors have one pair of field with charge $+1$ and $-1$. Dualizing this sector then yields $\UU(\tilde N)$ duals with an extra flavor associated with the gauged baryons. It would be interesting to work out the details of such dual phases.

 Another useful check of the dualities found here consists of engineering an RG flow connecting these dualities to the ones discussed in \cite{Kapustin:2011vz} in presence of CS terms. 
 In addition, real mass flows for the fundamentals would also induce chiral dualities, similar to the ones found in \cite{Benini:2011mf,Amariti:2020xqm,Amariti:2022lbw}.
 Other possible  dualities that one can construct in these cases   are inspired by the construction of 
 \cite{Benini:2017dud}, and require the presence of linear monopole superpotentials.
 Furthermore, starting from the pure 3d dualities obtained here, one may also consider real mass flows that give rise to CS terms for the tensors by appropriately turning on real masses along the $\UU(1)_X$ symmetry, finding dualities for $\UU(N)$ SQCD with non standard CS terms for the abelian sectors.

\section*{Acknowledgments}
This work  has been supported in part by the Italian Ministero dell'Istruzione, Università e Ricerca (MIUR), in part by Istituto Nazionale di Fisica Nucleare (INFN) through the “Gauge Theories, Strings, Supergravity” (GSS) research project.

\appendix

\section{The double scaling limit}
\label{dsrev}

In this appendix we review the basic aspects of the double scaling limit adopted here to reduce the superconformal index to the three sphere partition function.
We refer the reader to the general derivation in \cite{Amariti:2024bdd} and we report here only the formulae necessary to our analysis in the body of the paper.
We start by considering the gauge and flavor fugacities in the superconformal index as
\begin{equation}
    z_j = \ee^{2 \pi \mi u_j},  \quad v_k = \ee^{2\pi \mi m_k}.
\end{equation}
Then we  define a basis for the fugacities of each field  as,
\begin{equation}
    (pq)^{R_a/2} \prod_{k} v_i^{e_k^a} \equiv y_a \equiv \ee^{2\pi \mi \Delta_a},
\end{equation}
In this basis, the balancing condition translates into a constraint on the new variables 
$\Delta_a $.
In the  double scaling limit  the real masses are taken to be large with a $1/r$ scaling in the  $r \rightarrow 0$ limit of the radius  of the circle. 
Therefore, we parametrize the 4d fugacities as
\begin{equation}
    \label{paramds}
    u_i = \sigma_i^* + \sigma_i {r}\,,  \qquad \Delta_k = \mu_k^* + \mu_k {r}\,,
\end{equation}
where we assigned a fixed value $\sigma_i^*$ and $ \mu_k^*$ in addition to a term scaling at order $\mathcal{O}(r)$.
Depending  on choice of the fixed values in (\ref{paramds}) some mode coming from the matter and gauge fields can be massless in the KK tower or become completely massive.
Considering a single gauge and flavor holonomy and denoting  $k = \sigma^* + \mu^*$ and $x r = (\sigma + \mu)r$ we have
 \begin{equation}
    \label{eq:el_gamma_double}
    \Gamma_e(k + x r) \underset{r \to 0}{\sim}
    \begin{cases}
        \ee^{-\frac{\mi \pi (x - \omega) }{ 6 r_1 \omega_1\omega_2}} \; \Gamma_h(x) \quad \quad & k \in \mathbb{Z} \\
        \ee^{\mi \pi Q(k + x r)} & k \notin \mathbb{Z}\,,
    \end{cases}
\end{equation}
with
\begin{equation}
  \begin{aligned}
    Q(k + xr) = 
    - \frac{1}{\omega_1 \omega_2} &\Bigg(
    \frac{ k (2k - 1)(k - 1)}{6 r^2} + 
    \frac{(x - \omega)\left( 6k (k - 1) + 1 \right)}{6r} + \\ & +
    \frac{(2k - 1)(6x^2 + \omega_1^2 + \omega_2^2 + 3 \omega_1\omega_2 - 12x\omega )}{12} \Bigg) + \mathcal{O}(r)\,.
  \end{aligned}
\end{equation}
This is the formula that we have used in the paper to reduce the superconformal index to the three sphere partition function in the 
double scaling limit.

\section{The confining $\UU(3) \times \UU(3)$ quiver}
\label{app:0}

In this appendix, we study a 3d $\mathcal{N}=2$ confining duality that we encountered in the discussion in the body of the paper. The model 
has been used above in order to explicitly work out the presence of the singlets identified with the electric monopoles in the dual phase.
\begin{figure}
\begin{center}
  \includegraphics[width=6cm]{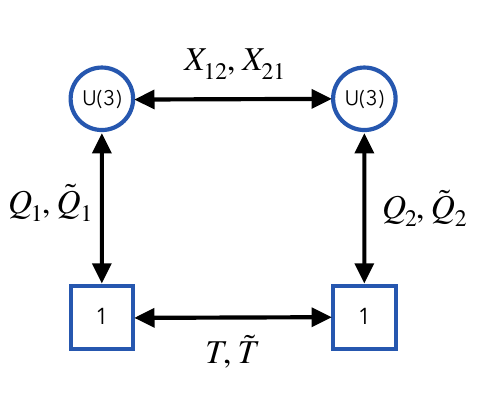}
  \end{center}
  \caption{A pictorial representation of the $\UU(3) \times \UU(3)$ studied in this appendix.}
    \label{Fig:quivU3}
\end{figure}
The 3d model considered in this appendix corresponds to the 
$\UU(3)_1\times \UU(3)_2$
quiver\footnote{The IR dynamics of the 4d  $\SU(N)\times \SU(N)$ parent of this model has been studied in \cite{Csaki:1997zg,Gremm:1997sz,Giveon:1997sn}. Here we restrict to the $N=3$ case and we gauge the baryonic symmetries. It should be interesting to study the 4d/3d reduction of this quiver, that admits a brane realization as well. }
 represented in Figure \ref{Fig:quivU3},  with a pair of bifundamental $X_{12}$ and $X_{21}$
and with one pairs of fundamentals $Q_i$ and antifundamentals  $\tilde Q_i$ 
for each $\UU(3)_i$ gauge factor. 
There is also a superpotential $W = (X_{12} X_{21})^2$. In the model studied in the body of the paper, we had also a set of flippers (denoted as $T,\tilde T, S_1,S_2,W_1,W_2$ below) that give rise to the following superpotential
\begin{equation}
W = (X_{12}X_{21})^2+T Q_1 X_{12} \tilde Q_2+\tilde T Q_2 X_{21} \tilde Q_1+\sum_{i=1}^2 (Q_i \tilde Q_i W_i+S_i Q_i \tilde Q_i X_{12} X_{21})\,.
\end{equation}

The next step consists of dualizing the $\UU(3)_1$ gauge factor. Using the rules of Aharony duality the dual superpotential is
\begin{eqnarray}
W &=& (Y_{12}Y_{21})^2  + K Y_{21} \tilde P_1 +  K P_1 Y_{12}+ T \tilde K  Q_2+\tilde T K \tilde Q_2+Q_2\tilde Q_2 W_2 +
\nonumber 
\\
&+&
S_1 K \tilde K+S_2 Q_2 \tilde Q_2 Y_{12} Y_{21} + V_1^+ v_1^- + V_1^- v_1^+ ,
\end{eqnarray}
where 
$K = X_{21} Q_1$ and $\tilde K = \tilde Q_1 X_{12}$, $Y_{12},Y_{2,1},P_1,\tilde P_1$ are dual to  $X_{12},X_{2,1},Q_1,\tilde Q_1$ 
and  the singlets $V_1^{\pm}$ flip the dual monopoles $v_1^{\pm}$ and  correspond to the electric monopoles.
The crucial aspect of this phase is that, if we start with two FI terms $\xi_1$ and $\xi_2$, the FI for this dual phase are $-\xi_1$ and $\xi_2-\xi_1$ respectively.

This last observation can be understood at the level of the three sphere partition function.
The identity between Aharony dual phases is
\begin{eqnarray}
\label{AharonyId}
&&
Z_{\UU(n)}^{(f \square;f \overline \square)} (\mu;\nu;\lambda)
=
Z_{\UU(f-n)}^{(f \square;f \overline \square)} (\omega-\nu;\omega-\mu;-\lambda) \;
c\left(\lambda  \sum_{a=1}^f (\mu_a - \nu_a)\right)  \nonumber \\
&& \times \,
\Gamma_h\Big(\pm \frac{\lambda}{2} + (f-n+1)\omega-\frac{1}{2} \sum_{a=1}^f(\mu_a+\nu_a)\Big)
\prod_{a,b=1}^f \Gamma_h(\mu_a+\nu_b)\,.
\end{eqnarray}

In the case at hand there are two $\UU(3)_1$ gauge factors and dualizing $\UU(3)_1$ forces us to treat $\UU(3)_2$ as a flavor symmetry group.
The partition function is
\begin{equation}
  \begin{aligned}
    &\Gamma_h(
      \tau_{\tilde S} +\mu^{(1)} +\mu^{(2)}
      )
    \Gamma_h(
      \tau_A +\nu^{(1)} +\nu^{(2)}
      )
    \prod_{j=1}^{2}
    \Gamma_h(
      \mu^{(j)}+ \nu^{(j)}; \mu^{(j)}+ \nu^{(j)} +\tau_A +\tau_{\tilde S}
      ) \\
    &\times\,
    \int \prod_{i=1}^3 \prod_{j=1}^{2} 
    \dd{\rho_i^{(j)}} 
    \ee^{\mi \pi \lambda_j \rho_{i}^{(j)}} 
    \Gamma_h\left(
      \tilde \mu^{(j)}\!-\!\rho_i^{(j)},\tilde \nu^{(j)}\!-\!\rho_i^{(j)}
      \right) \\
    &\times\,
    \frac{
      \prod_{i,\ell=1}^3 
      \Gamma_h\left(
        \tilde \tau_A + \rho_{\ell}^{(1)} + \rho_{i}^{(2)}, \tilde \tau_{\tilde S} - \rho_{\ell}^{(1)} - \rho_{i}^{(2)}
        \right)
    }{
      \prod_{1\leq i<\ell \leq 3}
      \prod_{j=1}^2
      \Gamma_h\left(
        \rho_{i}^{(j)}\!-\!\rho_{\ell}^{(j)}\right)
    }\,,
  \end{aligned}
\end{equation}
with the constraints
\begin{eqnarray}
&&
\tilde \mu^{(j)} +\tilde \nu^{(j)}=2\omega - \mu^{(j)} - \nu^{(j)}\,,
\nonumber \\
&&
\tilde\mu^{(1)}+\tilde\mu^{(2)} + \tilde\tau_{\tilde S} = 2 \omega-
\mu^{(1)}-\mu^{(2)} - \tau_{\tilde S}\,,
\nonumber \\
&&
\tilde\nu^{(1)}+\tilde\nu^{(2)} + \tilde\tau_{A} = 2 \omega-
\nu^{(1)}-\nu^{(2)} - \tau_{A}\,,
\nonumber \\
&&
\tau_{\tilde S}+\tau_A =\tilde  \tau_{\tilde S}+\tilde\tau_A =\omega\,.
\end{eqnarray}

We can then apply the rules of Aharony duality on $\UU(3)_1$ reducing it to $\UU(1)_1$. We observe that the $\UU(3)_2$ gauge theory is confining (it corresponds to a limiting case of Aharony duality). After confining this node and integrating out the massive fields, we are left with the $\UU(1)_1$ gauge node with one flavor. This theory confines as well. The flipper structure of the original $\UU(3)\times \UU(3)$ quiver is such that all the singlets in this chain of dualities are massive, except for the monopoles that arise after each duality. The final form of the partition function is
\begin{equation}
\label{finalU3U3}
\Gamma_h\left(
\pm \frac{\Lambda_2-\Lambda_1}{2} -\frac{1}{2} \sum_{\ell=1}^2(\mu^{(\ell)}+\nu^{(\ell)} )\right)
\prod_{\ell=1}^2
\Gamma_h\left(\pm \frac{\Lambda_\ell}{2} +\omega-\frac{1}{2} \left(\mu^{(\ell)}+\nu^{(\ell)} \right)\right),
\end{equation}
corresponding to the expected  three pairs of monopoles and anti-monopoles, charged under the two topological symmetries.
Observe that, in the body of the paper, we studied a case arising from a model with an orientifold that forces the two FI terms to be equal.

\section{$\UU(4)$ with an antisymmetric and two fundamental flavors}
\label{app:U4}
In the body of the paper, we exploited  a confining duality involving 
$\UU(4)$ with a pair of conjugated antisymmetric $A$ and $\tilde A$ and two pairs of fundamental flavors $Q_{1,2}$ and $\tilde Q_{1,2}$.

In this appendix, we prove the result explicit at the level of the field theory and of 
the three sphere partition function.

The starting point of the proof is a confining duality studied originally in \cite{Csaki:2014cwa} and further discussed in \cite{Nii:2019ebv} involving $\SU(4)$ with two antisymmetrics $A_{1,2}$ and  two pairs of fundamental flavors $Q_{1,2}$ and $\tilde Q_{1,2}$.
This duality is obtained through ARSW reduction of a 4d confining duality with 
two antisymmetrics and three pairs of fundamental flavors.
Observe that, in the case of $\SU(4)$, the antisymmetric is self-conjugated, and thus it is not necessary to distinguish between the antisymmetric and the conjugated antisymmetric. This gives origin to a further $\SU(2)$ flavor symmetry rotating the two antisymmetric, which is broken when the baryonic $\UU(1)$ is gauged. Indeed, the gauging of the baryonic symmetry involves the non abelian $J_3$ generator of such an $\SU(2)$.

 The $\SU(4)$ confining duality maps the $\SU(4)$ theory to a WZ model, where the singlets (using the conventions of \cite{Nii:2019ebv}) are 
 $M_0= Q \tilde Q$, $M_{2} = Q A^2 \tilde Q$ (where the contraction between the two antisymmetric is a singlet of the $\SU(2)$ flavor symmetry), $B=A Q^2$,
 $\tilde B = A \tilde Q^2$ and $T=A^2$ (in the adjoint of the $\SU(2)$ flavor symmetry).
 The superpotential of the dual phase is 
 \begin{equation}
 W = Y_{\SU(2)}^{(\text{bare})} (T^2 \det M_0 + \det M_2 + T B \tilde B)+
Y_{\SU(2)\times \SU(2)}^{(\text{bare})}  (M_0 M_2 +B \tilde B)\,.
 \end{equation}

In the body of the paper we have used a flipped version of such duality, where
the singlets $M_0$, $M_2$ and some components of  $B$ and $\tilde B$ (corresponding to $A_2 Q^2$ and $A_1 \tilde Q^2$) have been set to zero in the chiral ring. Furthermore, the presence of an electric quartic superpotential involving only the two antisymmetrics sets $T=0$ in the chiral ring.
The electric superpotential is as follows
\begin{equation}
W = (A \tilde A)^2 + N_2 Q (A \tilde A) \tilde Q + N_0 Q \tilde Q + \tilde R A \tilde Q^2
+  R \tilde A  Q^2\,.
\end{equation}
Then, by gauging the baryonic symmetry, we have a $\UU(1)$ dual model with superpotential 
 \begin{equation}
 W = Y_1 b \tilde b\,,
 \end{equation}
where $b$ and $\tilde b$ correspond to the combinations $A Q^2$ and $\tilde A \tilde Q^2$ and $Y_1$ is the $ Y_{\SU(2)\times \SU(2)}^{(\text{bare})} $ monopole.
Observe that, there is a second monopole $Y_0$ corresponding to $Y_{\SU(2)}^{(\text{bare})}$ that is not interacting anymore but is left in the low energy spectrum.
The $\UU(1)$ dual model can be further dualized to a pair of monopoles and a meson. The meson is massive together with $Y_1$, while the other two monopoles remain in the low energy spectrum.
The confining duality of the flipped $\UU(4)$ consists then of three monopoles, two charged under the topological symmetry while one uncharged.

In the following, we will reproduce this behavior by gauging the  baryonic $\UU(1)$ symmetry on the identity between the three sphere partition functions.
Let us start from the $\SU(4)$ identity:
\begin{equation}
  \label{csakiNii}
  \begin{aligned}
    Z_{\SU(4)}^{(2\square, 2 \overline \square, 2A)} &=
    \frac{1}{4!} \int 
    \prod_{i=1}^4  \dd{\sigma_i}
    \delta \left(\sum_{i=1}^4 \sigma_i\right)
    \frac{
      \prod_{i<j} 
      \Gamma_h(
        \sigma_i +\sigma_j + \tau_{A_1}
        )
      \Gamma_h(
        -\sigma_i -\sigma_j + \tau_{A_2}
        )
    }{
      \prod_{i<j} 
      \Gamma_h(
        \pm(\sigma_i + \sigma_j)
        )
      } \\
    &\times \prod_{a=1}^2 \prod_{i=1}^4
    \Gamma_h(
      \mu_a-\sigma_i,\nu_a+\sigma_i
      ) = \\
    &=
    \prod_{1\leq a< b\leq 2}
    \Gamma_h(
      \mu_a\!+\!\nu_b,\mu_a\!+\!\nu_b\!+\!\tau_{A_1}\!+\!\tau_{A_2}
      )
    \prod_{\ell=1}^2
    \Gamma_h(
      \tau_{A_\ell}\!+\!\mu_1\!+\!\mu_2,\tau_{A_\ell}\!+\!\nu_1 +\nu_2
      ) 
    \\ &\times \,
    \prod_{1\leq \ell\leq k\leq 2} 
    \Gamma_h(
      \tau_{A_\ell}+ \tau_{A_k}
      )
    \prod_{j=1}^2 
    \Gamma_h(
      2\omega-\sum_{a=1}^{2} (\mu_a + \nu_a)- j(\tau_{A_1}+\tau_{A_2})
      ).
  \end{aligned}
\end{equation}
Even if this identity, at least to our knowledge, is not explicitly written in the literature, it can be derived reducing the 4d identity between the superconformal indices and then by considering the real mass flow. The derivation is straightforward and we leave it to the interested reader.

Observe that, the second antisymmetric has been written using negative signs for the $\sigma$'s because in this way the baryonic symmetry can be gauged giving rise to the expected conjugated field content.

We can then gauge a baryonic $\UU(1)$ symmetry adding an integral $\int \dd{\mathbf{b}} \ee^{2\pi \mi \lambda (4 \mathbf{b})}$ and Fourier transform the delta-function $\int \dd{\xi} \ee^{2 \pi \mi \xi \sum_{i=1}^{4} \sigma_i}$. If we shift the integration variables as $\sigma_i \rightarrow \sigma_{i} -\mathbf{b}$ and integrate over $\dd{\mathbf{b}}$, we obtain in the integrand a delta function $\delta(\xi - \lambda)$.

The electric side of \eqref{csakiNii} becomes 
\begin{equation}
\frac{1}{4 \times 4!} \int 
\prod_{i=1}^4  \dd{\sigma_i}
\frac{
\prod_{i<j} \Gamma_h(\sigma_i +\sigma_j +\hat \tau_{A_1})\Gamma_h(-\sigma_i -\sigma_j + \hat \tau_{A_2}) }{\prod_{i<j} \Gamma_h(\pm(\sigma_i + \sigma_j))}
\prod_{a=1}^2\prod_{i=1}^4\Gamma_h(\hat \mu_a-\sigma_i,\hat \nu_a+\sigma_i),
\end{equation}
where the hatted mass parameter differ from the ones above by the absence of a baryonic symmetry.
Furthermore, we consider the superpotential $W \propto (A_1 A_2)^2$ forcing $\hat \tau_{A_1} + \hat \tau_{A_2} = \omega$.

On the dual side, we can add the integral over the baryonic symmetry and consider the constraint from the quartic superpotential.
We obtain
\begin{equation}
  \label{dualu4}
  \begin{aligned}
    &\Gamma_h(
      \tau_{A_1}\!+\!\mu_1 \!+\!\mu_2,\tau_{A_2}\!+\!\nu_1 \!+\!\nu_2
      )
    \prod_{a,b=1,2} \Gamma_h(
      \mu_a \!+\!\nu_b,\mu_a \!+\!\nu_b\!+\!\omega
      )
    \prod_{j=0}^1 \Gamma_h\left(
      j \omega\!-\!\sum_{a=1}^{2} (\mu_a\! +\! \nu_a)
      \right) \\
    & \times \frac{1}{4} 
    \int \dd{(4\mathbf{b})} 
    \ee^{2 \pi \mi \lambda (4\mathbf{ b})} \Gamma_h(
      \hat \tau_{A_2}+\hat \mu_1 +\hat \mu_2+4  \mathbf{b},\hat \tau_{A_1}+\hat\nu_1 +\hat\nu_2-4  \mathbf{b}
      )\,,
  \end{aligned}
\end{equation}
that corresponds to the $\UU(1)$ SQED  expected from the field theoretical discussion.
Observe that, the singlets in the first line of \eqref{dualu4} are flipped, and on the final identity used in the body of the paper they appear on the LHS indeed.

We can conclude by performing the $\UU(1)$ integral, obtaining one meson that simplifies against the contribution of the monopole with $j=1$ in the second line of \eqref{dualu4} and two monopoles that contribute as 
\begin{eqnarray}
\Gamma_h\left(\pm \frac{\lambda}{2} + \frac{\omega}{2}-\frac{1}{2}(\mu_1 +\mu_2+\nu_1 +\nu_2)\right).
\end{eqnarray}

For completeness we rewrite the final formula that we have used in the body of the paper as follows
\begin{equation}
  \label{finalconfininjg}
  \begin{aligned}
    \prod_{a,b=1,2}&\Gamma_h(
      2\omega\!-\!\mu_a\!-\!\nu_b,\omega\!-\!\mu_a\!-\!\nu_b
      ) 
    \Gamma_h(
      \tau_{A}\!+\!\mu_1\!+\!\mu_2,\tilde \tau_{A}\!+\!\nu_1\!+\!\nu_2
      ) 
    Z_{\UU(4)}^{(2\square, 2 \overline \square, A;\tilde A)}
    (\vec \mu;\vec \nu;\tau;\tilde \tau)
    \\ =\;&
    \Gamma_h\left(
      -\sum_{a=1}^{2} (\mu_a + \nu_a)
      \right)
    \Gamma_h\left(
      \pm \frac{\lambda}{2} + \frac{\omega}{2}-\frac{1}{2}\sum_{a=1}^{2} (\mu_a + \nu_a)
      \right),
  \end{aligned}
\end{equation}
with $\tau+\tilde \tau=\omega$.

\section{Remarks on monopole operators}
\label{app:mono}
In the body of the paper, we encountered monopole operators arising as singlets in the dual phases with superpotential interactions
setting to zero the dual monopoles, which is typical of 3d Aharony-like dualities. 
The monopoles have been referred to as $V_{\pm}$ and $V_0$ in all the examples. The subscript refers to their topological charge.
In this appendix, we will further discuss such monopoles (we refer the reader to \cite{Borokhov:2002ib,Borokhov:2002cg,Kapustin:2005py,Intriligator:2013lca,Aharony:2013dha,Csaki:2014cwa,Amariti:2015kha,Nii:2019ebv,Benvenuti:2020wpc,Amariti:2024gco} for a comprehensive discussion).
We start our discussion by considering an $\SU(N)$ case in presence of two-index tensors (symmetric or antisymmetric)  and we look for the monopole with minimal GNO flux. Such monopoles have been studied in great detail in \cite{Nii:2019ebv} for models with antisymmetric matter, and they correspond to monopole that induces an $\SU(N-2) \times \UU(1)_1 \times \UU(1)_2$ gauge  
symmetry breaking pattern
where
\begin{equation}
	\begin{split}
		& U(1)_1 \sim \text{diag}(1, \underbrace{0,...,0}_{N-2},-1), \\
		& U(1)_2 \sim \text{diag} (N-2,\underbrace{-2,...,-2}_{N-2}, N-2).
	\end{split}
\end{equation}
The branching rules for this breaking are
\begin{align}
		& \square \longrightarrow 	\square_{(0,-2)} \oplus \textbf{1}_{(1,N-2)} \oplus \textbf{1}_{(-1,N-2)} \nonumber \\
		& \overline{\square} \longrightarrow 	\overline{\square}_{(0,2)} \oplus \textbf{1}_{(-1,-(N-2))} \oplus \textbf{1}_{(1,-(N-2))} \nonumber \\
		& \symmF \longrightarrow \symmF_{(0,-4)} 	\oplus \square_{(1,N-4)} \oplus \square_{(-1,N-4)} \oplus \textbf{1}_{(2,2N-4)} \oplus \textbf{1}_{(-2,2N-4)} \oplus \textbf{1}_{(0,2N-4)} \nonumber \\
		& \symmBF \longrightarrow  \symmBF_{(0,4)} \oplus \overline{\square}_{(-1,-(N-4))} \oplus \overline{\square}_{(1,-(N-4))} \oplus \textbf{1}_{(-2,-(2N-4))} \oplus \textbf{1}_{(2,-(2N-4))} \nonumber \\ & \phantom{\symmBF \longrightarrow} \oplus \textbf{1}_{(0,-(2N-4))} \nonumber \\
		& \asymmF \longrightarrow \asymmF_{(0,-4)} 	\oplus \square_{(1,N-4)} \oplus \square_{(-1,N-4)} \oplus \textbf{1}_{(0,2N-4)} \nonumber \\
		& \asymmBF \longrightarrow 	\asymmBF_{(0,4)} \oplus \overline{\square}_{(-1,-(N-4))} \oplus \overline{\square}_{(1,-(N-4))} \oplus \textbf{1}_{(0,-(2N-4))} \nonumber \\
		& \textbf{adj} \longrightarrow \textbf{adj}_{(0,0)} \oplus \overline{\square}_{(1,N)} \oplus \overline{\square}_{(-1,N)} \oplus \square_{(1,-N)} \oplus \square_{(-1,-N)} \oplus \textbf{1}_{(0,0)} \oplus \textbf{1}_{(0,0)} \oplus \textbf{1}_{(2,0)} \nonumber \\ & \phantom{\textbf{adj} \longrightarrow} \oplus \textbf{1}_{(-2,0)} \nonumber \\
	\end{align}
The Coulomb branch corresponds to the generator $\UU(1)_1$ and the bare monopole is built by dualizing its photon.
For this reason, they are denoted as $Y_{\SU(N-2)}^{(\text{bare})}$. The field content contributing to the monopole charge has to be read  after applying the branching rules for the symmetry breaking pattern above. Indeed, the massive fields (the components charged under $\UU(1)_1$) are massive on the Coulomb branch and are integrated out. The remaining massless fields are precisely those that must be considered in order to determine the monopole charges.

First, one needs to consider the gauge charge of the bare Coulomb branch  $Y_{\SU(N-2)}^{(\text{bare})}$.
For example, when the number of fundamental and antifundamental fields is different, the bare Coulomb branch operators are not gauge invariant, because  $k_{\UU(1)_1,\UU(1)_2} \neq 0$.
Here, considering the breaking pattern above, we have
\begin{equation}
  k_{\UU(1)_1,\UU(1)_2} = 
  \frac{N-2}{2}
  \biggl(
    \left(
      F-\bar{F}
    \right)  + 
    (N-4) \left(
      A-\bar{A}
    \right)
     + 
    N \left(
      S-\bar{S}
    \right)
  \biggr)
\end{equation}
In the examples discussed in the body of the paper, $k_{\UU(1)_1,\UU(1)_2} = 0$, i.e. the bare $\SU(N)$  monopole is gauge invariant. In addition, its baryonic charge is zero and such a monopole survives the gauging of $\UU(1)_B$. Indeed, it corresponds to the monopole $V_0$ in the $\UU(N)$ dualities. 

Furthermore, after gauging the baryonic symmetry we have additionally introduced a topological symmetry $\UU(1)_J$ and we can define monopole operators charged under such symmetry. 
These are the  monopoles $V_{\pm}$ discussed in the body of the paper, they have flux $\pm 1$ and correspond to the Coulomb branches $\text{diag}(1,0,...,0)$ and $\text{diag}(0,0,...,-1)$.
They are gauge invariant and appear  in all the pure 3d dualities \eqref{AAtpure3d}, \eqref{SStpure3d} and  \eqref{chiralpure3d} found here.

\bibliographystyle{JHEP}
\bibliography{refBS.bib}
\end{document}